\documentclass[aps,prd,twocolumn,longbibliography,notitlepage]{revtex4-1}
\usepackage{amsmath,amssymb,bbm,bm}
\usepackage[usenames,dvipsnames]{xcolor}
\usepackage{hyperref}
\usepackage{graphicx}
\usepackage{braket}
\usepackage{maybemath}

\hypersetup{pdfauthor={},pdftitle={},%
colorlinks, linktocpage=true, pdfstartpage=3, pdfstartview=FitV,%
urlcolor=orange, linkcolor=blue, citecolor=red }

\definecolor{garrosgreen}{rgb}{0.1, 0.4, 0.1}
\definecolor{dartmouthgreen}{rgb}{0.05, 0.5, 0.06}
\definecolor{duelferred}{rgb}{0.7, 0.2, 0.1}
\definecolor{cambridgeblue}{rgb}{0.1, 0.3, 1.0}
\definecolor{oxfordblue}{rgb}{0.05, 0.2, 0.7}

\newcommand{\comment}[1]{}

\newcommand{\bew}{\begin{widetext}}
\newcommand{\eew}{\end{widetext}}

\newcommand{\dd}{{\mathrm{d}}}
\newcommand{\ii}{{\mathrm{i}}}
\newcommand{\ee}{{\mathrm{e}}}

\newcommand{\calN}{{\mathcal{N}}}
\newcommand{\calO}{{\mathcal{O}}}

\newcommand{\stdrule}{\rule[-3mm]{0mm}{8mm}}

\begin{document}
\title{Correlation Functions of the Anharmonic Oscillator:\\
Numerical Verification of Two--Loop Corrections to the Large--Order Behavior}

\author{Ludovico T. Giorgini}

\affiliation{Nordita, Royal Institute of Technology and Stockholm University,
Stockholm 106 91, Sweden}

\author{Ulrich D. Jentschura}

\affiliation{Department of Physics, Missouri University of Science and
Technology, Rolla, Missouri 65409, USA}

\affiliation{MTA-DE Particle Physics Research Group, P.O. Box 51, H-4001
Debrecen, Hungary}

\author{Enrico M. Malatesta}

\affiliation{Artificial Intelligence Lab, Bocconi University, 20136 Milano, Italy}

\author{Giorgio Parisi}

\affiliation{Dipartimento di Fisica, Sapienza Universit\`a di Roma, P.le Aldo
Moro 5, 00185 Rome, Italy}

\affiliation{Istituto Nazionale di Fisica Nucleare, Sezione di Roma I, P.le A.
Moro 5, 00185 Rome, Italy}

\affiliation{Institute of Nanotechnology (NANOTEC) - CNR, Rome unit, P.le A.
Moro 5, 00185 Rome, Italy}

\author{Tommaso Rizzo}

\affiliation{Institute of Complex Systems (ISC) - CNR, Rome unit, P.le A. Moro
5, 00185 Rome, Italy}

\affiliation{Dipartimento di Fisica, Sapienza Universit\`a di Roma, P.le Aldo
Moro 5, 00185 Rome, Italy}

\author{Jean Zinn-Justin}

\affiliation{IRFU/CEA, Paris-Saclay, 91191 Gif-sur-Yvette Cedex, France}

\begin{abstract}
Recently, the large-order behavior of correlation
functions of the $O(N)$-anharmonic oscillator 
has been analyzed by us in [L. T. Giorgini {\em et el.},
Phys.~Rev.~D {\bf 101}, 125001 (2020)].
Two-loop corrections about the instanton 
configurations were obtained for the 
partition function, and the two-point and four-point functions, 
and the derivative of the two-point function at zero momentum 
transfer. Here, we attempt to verify the 
obtained analytic results against numerical calculations
of higher-order coefficients 
for the $O(1)$, $O(2)$, and $O(3)$ oscillators,
and demonstrate the drastic improvement of the 
agreement of the large-order asymptotic estimates
and perturbation theory upon the inclusion of the 
two-loop corrections to the large-order behavior.
\end{abstract}
\maketitle

\clearpage 
\tableofcontents{}

%
%
\section{Introduction}
\label{sec1}

For a long time, it has been a dream of physics research 
to overcome the predictive limits of perturbative 
quantum field
theory. Typically, Feynman diagram calculations
become more computationally expensive in large loop orders,
to the point where diminishing returns~\cite{GiEtAl2021outreach}
upon the addition of yet another loop 
limit the predictive power of perturbation theory,
and of Feynman diagram calculations.
In order to overcome these limits,
analytic techniques have been developed 
over the past decades, to analyze the large-order 
behavior from the complementary limit 
of ``infinite-loop-order'' Feynman 
diagrams~\cite{BrPaZJ1977,BrPa1978}.
These techniques are based on various nontrivial
observations. The first is that,
upon an analytic continuation of the 
coupling constant of a theory into a physically 
``unstable'' domain~\cite{Dy1952} where partition functions
acquire an imaginary part,
one can write dispersion relations
which relate the behavior of the theory 
for a coupling constant small in absolute magnitude,
within in the unstable domain, to the 
large-order behavior of perturbation theory
(equivalent to the ``infinite-order Feynman diagrams'').
The imaginary part of the partition functions, and of the 
correlation functions, is related to so-called 
instanton configurations~\cite{JeZJ2004plb,ZJJe2004i,ZJJe2004ii}.
The second observation is that perturbations
about the instanton configurations
can be mapped onto corrections to the large-order 
behavior of perturbation theory, thus making the
theory amenable to a more accurate analysis
in the domain of large loop orders.
The latter perturbative calculations in the 
instanton sector are related to an expansion
of perturbative coefficients in powers of inverse loop orders
$1/K$, where $K$ denotes the loop order.
The leading term, of course, 
as is well known, describes the factorial divergence 
of perturbation theory in large orders
of the coupling constant~\cite{BeWu1969,BeWu1971,BeWu1973,%
BrPaZJ1977,BrLGZJ1977prd1,BrLGZJ1977prd2}.

Previously, the calculation of corrections about the 
large-order behavior was reported for the 
partition function of anharmonic oscillators~\cite{JeZJ2011, MaPaRi2017}.
Recently~\cite{GiEtAl2020}, corrections to the large-order
behavior of perturbation theory have been 
obtained for the two-point and four-point functions
of the $O(N)$ quartic anharmonic oscillator.
Also, the partition function of the 
$O(N)$ oscillator was studied, 
and results were obtained for the derivative of the 
two-point function at zero momentum transfer~\cite{GiEtAl2020}.
These results, however, have not been compared yet
to an explicit calculation of perturbative coefficients
for the respective functions in large orders.
In this work, in order to make the comparison 
possible, we derive general expressions for 
the perturbative coefficients of the correlation
functions at zero momentum transfer of the one-, two- and three-dimensional
isotropic anharmonic oscillator.

In this context, it is extremely interesting to 
investigate the ``rate of convergence of the 
expansion about infinite loop order'', 
i.e., to investigate to which extent the calculation of the 
corrections of order $1/K$ to the leading factorial 
asymptotics improves the agreement of low-order
perturbation theory (corresponding to the 
successive perturbative evaluation of loops).
In this paper, we thus analyze the perturbation series of the 
one-dimensional $O(N)$
isotropic quantum harmonic oscillator with a quartic perturbation,
which is otherwise referred to as the $N$-vector model.
Higher orders of perturbation theory are calculated 
for the two-point function, the 
four-point function, and the correlator 
with a wigglet insertion, 
and compared to the results recently reported
in Ref.~\cite{GiEtAl2020}. Our Hamiltonian is therefore
\begin{equation}
H = - \frac{1}{2} \frac{\partial^2}{\partial \vec q^{\,2}} + 
\frac12 \, \vec q^{\,2} + \frac{g}{4} \vec q^{\,4} \,,
\qquad
\vec q \equiv \sum_{i=1}^N q_i \, \hat{\mathrm{e}}_i \,,
\end{equation}
where $g$ is the coupling constant.  The rest of the paper is organized as
follows. In Sec.~\ref{sec2}, we start by analyzing the simple $N=1$ case,
whereas in Sec.~\ref{sec3}, we discuss the general $N$-dimensional case. In
Our results are compared with those 
of Ref.~\cite{GiEtAl2020} and Sec.~\ref{sec4}.
Conclusions are reserved for Sec.~\ref{sec5}.

%
%
\section{One--Dimensional Quantum Anharmonic Oscillator}
\label{sec2}


We start with a discussion of the method of calculation for the 
perturbative corrections to the correlation functions 
exposed in the previous section. The case with a trivial internal symmetry 
group is the easiest ($N=1$).
Since when $N=1$ the unperturbed Hamiltonian is non-degenerate,
we can use standard non-degenerate Rayleigh-Schr\"odinger 
perturbative theory techniques. 
We can write the 
perturbative expansions as follows,
\begin{subequations}
\label{unperturbed}
\begin{align}
H^{(0)} =& \; - \frac{1}{2} \frac{\partial^2}{\partial q^2} + 
\frac12 \, q^2 \,,
\qquad 
\delta H =  \frac{g}{4} \, q^4 \,,
\\[0.1133ex]
| E_n \rangle =& \; | E^{(0)}_n \rangle + 
| E^{(1)}_n \rangle + | E^{(2)}_n \rangle + \dots \,,
\\[0.1133ex]
| \psi_n \rangle =& \; | \psi^{(0)}_n \rangle + 
| \psi^{(1)}_n \rangle + | \psi^{(2)}_n \rangle + \dots \,.
\end{align}
\end{subequations}
The unperturbed problem, with the 
unperturbed Hamiltonian $H^{(0)}$, is solved
by the unperturbed states $| \psi^{(0)}_n \rangle$,
\begin{equation}
H^{(0)} \; | \psi^{(0)}_n \rangle =
E^{(0)}_n \; | \psi^{(0)}_n \rangle \,,
\end{equation}
leading to the unperturbed energy 
eigenvalues $E^{(0)}_n$.
The perturbative corrections 
$E^{(K=1,2,3,\dots)}_n$ are each proportional to 
$g^K$ and describe a perturbation series in $g$.
On the basis of a well-known recursive scheme,
we can calculate higher-order (in $K$) 
perturbations to the wave functions, as follows:
\begin{subequations}
\label{recursive}
\begin{align}
g^0: & \; ( H^{(0)} - E^{(0)}_n ) \, | \psi^{(0)}_n \rangle = 0 \,,
\\
g^1: & \; (  H^{(0)} - E^{(0)}_n ) \, | \psi^{(1)}_n \rangle = 
(  E^{(1)}_n - \delta H ) \, | \psi^{(0)}_n \rangle = 0 \,,
\\
g^2: & \; (  H^{(0)} - E^{(0)}_n ) \, | \psi^{(2)}_n \rangle =
(  E^{(1)}_n - \delta H ) \, | \psi^{(1)}_n \rangle 
\nonumber\\ 
& \; \qquad + E^{(2)}_n \, | \psi^{(0)}_n \rangle \,,
\\
g^3: & \; (  H^{(0)} - E^{(0)}_n ) \, | \psi^{(3)}_n \rangle =
(  E^{(1)}_n - \delta H ) \, | \psi^{(2)}_n \rangle
\nonumber\\
& \; \qquad + E^{(2)}_n \, | \psi^{(1)}_n \rangle
+ E^{(3)}_n \, | \psi^{(0)}_n \rangle \,,
\\
g^K: & \; (  H^{(0)} - E^{(0)}_n ) \, | \psi^{(K)}_n \rangle =
(  E^{(1)}_n - \delta H ) \, | \psi^{(K-1)}_n \rangle
\nonumber\\  
& \; 
+ E^{(2)}_n | \psi^{(K-2)}_n \rangle
+ E^{(3)}_n | \psi^{(K-3)}_n \rangle 
+ \ldots 
+ E^{(K)}_n | \psi^{(0}_n \rangle \,.
\end{align}
\end{subequations}
This recursive algorithm allows us to calculate
the $K$th-order perturbation $| \psi^{(K)}_n \rangle$
to the wave function.
Note that the calculation of the $K$th-order
energy perturbation $E^{(K)}_n$ only 
requires the wave function to order $K-1$, 
\begin{equation}
E^{(K)}_n = \langle \psi^{(0)}_n | \delta H | \psi^{(K-1)}_n \rangle \,.
\end{equation}
The wave function perturbations $| \psi^{(K)}_n \rangle$
are orthogonal to the unperturbed 
state $| \psi^{(0)}_n \rangle$,
\begin{equation}
\langle \psi^{(0)}_n | \psi^{(1)}_n \rangle = 
\langle \psi^{(0)}_n | \psi^{(2)}_n \rangle = 
\langle \psi^{(0)}_n | \psi^{(3)}_n \rangle = \ldots = 0 \,.
\end{equation}

In order to solve the recursive scheme
given by Eq.~\eqref{recursive},
for a given unperturbed reference state $| \psi^{(0)}_n \rangle$,
we can define the reduced Green function
of the unperturbed problem,
\begin{equation}
\hat{T} = \left( \frac{1}{H^{(0)} - E^{(0)}_n} \right)' \,,
\end{equation}
where the inverse is taken over the Hilbert space
of unperturbed states 
orthogonal to the reference state $| \psi^{(0)}_n \rangle$.
The reduced Green function is sometimes 
denoted as $G'$ in the literature, but we avoid the notation
here because the symbol $G$ is already used extensively 
in other parts of our considerations, in order
to denote perturbative coefficients. 
$\hat{T}$ has the following matrix elements in 
the basis of unperturbed eigenstates,
\begin{equation}
\label{direct}
\hat{T}_{m,m'} = 
\left< \psi^{(0)}_m \left| \left( \frac{1}{H^{(0)} - E^{(0)}_n} \right)' 
\right| \psi^{(0)}_{m'} \right> =
\frac{\delta_{m,m'}}{E_m^{(0)} - E_n^{(0)}} \,, 
\end{equation}
assuming that $m \neq n$ and $m' \neq n$.
Furthermore, we have $\hat{T}_{n,n}=0$
for $m = m' = n$. We can write the perturbation 
$\delta H = g \, q^4/4$ in the unperturbed basis
simply by using the representation of the position operator 
in terms of the creation and annihilation operators
$a^\dagger$ and $a$ of the unperturbed Hamiltonian,
\begin{equation}
q = \frac{1}{\sqrt{2}} \left(a+ a^\dagger \right) \,.
\end{equation}
We recall that the lowering and raising 
operators $a$ and $a^\dagger$ act on the 
unperturbed state $| \psi^{(0)}_n \rangle$ as
follows,
\begin{subequations}
\begin{align}
a | \psi^{(0)}_n \rangle =& \; 
\sqrt{n} \; | \psi^{(0)}_{n-1} \rangle \,, \\
a^\dagger | \psi^{(0)}_n \rangle =& \;
\sqrt{n+1} \; | \psi^{(0)}_{n+1} \rangle \,.
\end{align}
\end{subequations}

From now on, we will switch to a notation where
\begin{equation}
| n \rangle \equiv | \psi_n \rangle 
\end{equation}
denotes the (perturbed) eigenstate of the full problem,
which includes the quartic term. 
Averages of a one-dimensional scalar theory can be interpreted as 
path-integral expressions which in turn can be written as summations over
eigenvalues of the corresponding quantum Hamiltonian. 
The two-point function has the translational 
invariance property
\begin{equation}
C^{(2)}(t_1, t_2) = 
C^{(2)}(0, t_2 - t_1) \equiv 
C^{(2)}(t_2 - t_1) \,,
\end{equation}
where the latter expression defines the 
correlation function $C^{(2)}(t_2 - t_1)$ 
of a single argument.
The correlation function can be expressed as follows,
\begin{equation}
\label{C2eq}
C^{(2)}(t)\equiv \langle q(0) \, q(t) \rangle
= \sum_{n>0}|\langle 0|q|n\rangle|^2 \, \ee^{-(E_n-E_0)|t|} \,,
\end{equation}
where $E_0$ is the perturbed energy of the ground state
(the ``vacuum''), and $E_n$ is the perturbed
energy of the state $| n \rangle$.
We have for its integrals
\begin{align} 
\int_{-\infty}^{+\infty} \dd t \, C^{(2)}(t) =& \;
2 \sum_{n>0} \frac{ |\langle 0|q|n\rangle|^2}{E_n-E_0} \,, 
\nonumber\\[0.1133ex]
\sim & \; \sum_{K=0}^\infty [ G^{(2)}(1,1) ]_K \, g^K \,,
\\[0.1133ex]
\int_{-\infty}^{+\infty} \dd t \, t^2\,C^{(2)}(t) =& \;
4 \sum_{n>0}
\frac{ |\langle 0|q|n\rangle|^2}{(E_n-E_0)^3} 
\nonumber\\[0.1133ex]
\sim & \; \sum_{K=0}^\infty [ G^{(\partial p)}(1,1) ]_K \, g^K \,.
\end{align} 
Here, the perturbative coefficients 
$[ G^{(2)}(N=1,1) ]_K$ and 
$[ G^{(\partial p)}(N=1,1) ]_K$ define an
asymptotic series in the coupling $g$ which 
exhibits well-known factorial growth for 
higher order in $K$. In the notation, we follow the conventions
of Ref.~\cite{GiEtAl2020}; some low-order coefficients
$[ G^{(2)}(1,1) ]_K$ and $[ G^{(\partial p)}(1,1) ]_K$ are
summarized in Table~\ref{table1}. A remark is in order.
The expression for the integral
$\int_{-\infty}^{+\infty} C^{(2)}(t) \, \dd t$ deceptively looks like 
the second-order expression for the energy shift 
due to a perturbative Hamiltonian proportional to $q$.
We should recall, though, that the 
the virtual-state eigenfunction $\ket{n}$ and
the ground-state eigenfunction $\ket{0}$,
as well as the energies $E_n$ and $E_0$, 
are the exact eigenfunctions and eigenenergies 
of the perturbed anharmonic oscillator and thus
contain contributions of arbitrarily higher orders
of perturbations proportional to $g^K$.

As a side remark, let us briefly consider the limit 
$g \to 0$, which describes the transition to the 
unperturbed harmonic oscillator. 
In this limit, one has

\begin{multline}
C^{(2)}(t) = \sum_{n>0}|\langle 0|q|n\rangle|^2 \, \ee^{-(E_n-E_0)|t|} 
\\
\to 
C^{(2)}(t) 
= \sum_{n>0}|\langle \psi^{(0)}_0 |q| \psi^{(0)}_n 
\rangle|^2 \, \ee^{-(E^{(0)}_n-E^{(0)}_0)|t|} 
\\
= |\langle \psi^{(0)}_0 |q| \psi^{(0)}_1 \rangle|^2 \, \ee^{-|t|} 
= \frac12 \, \ee^{-|t|} \,,
\end{multline}
which is the well-known two-point correlation function 
for the unperturbed one-dimensional theory~\cite{ZJ2005}.

\color{black}

The connected four-point function is defined as
\begin{multline}
C(t_1,t_2,t_3,t_4)\equiv 
\langle q(t_1) \, q(t_2) \, q(t_3) \, q(t_4) \rangle \\
- C^{(2)}(t_1-t_2)C^{(2)}(t_3-t_4) -C^{(2)}(t_3-t_2)C^{(2)}(t_1-t_4) \\
-C^{(2)}(t_1-t_3)C^{(2)}(t_2-t_4) \,.
\end{multline}
For $0=t_1<t_2<t_3<t_4$, we can write 
\begin{multline}
\label{connect4}
C^{(4)}(t_2,t_2,t_4) \equiv \langle q(0)q(t_2)q(t_3)q(t_4) \rangle \\
= \langle 0 | q\, \ee^{-(H-E_0)\Delta_3}\,
q\,\ee^{-(H-E_0)\Delta_2}\, 
q\, \ee^{-(H-E_0)\Delta_1} q |0\rangle \\
- C^{(2)}(\Delta_1) \, C^{(2)}(\Delta_3) 
- C^{(2)}(\Delta_1+\Delta_2+\Delta_3) \, C^{(2)}(\Delta_2) \\
- C^{(2)}(\Delta_1+\Delta_2) \, C^{(2)}(\Delta_2+\Delta_3) \,,
\end{multline}
where $\Delta_i\equiv t_{i+1}-t_i$. 
We have chosen $t_1 = 0$. 
For $t_2$, we have two equivalent regions
($t_2$ can be either positive or negative),
while for $t_3$, we have three equivalent regions.
Finally, for $t_4$, one encounters four equivalent regions,
bringing the number of equivalent integration
regions to $4 \times 3 \times 2 = 24$.
The selected integration region 
$0=t_1<t_2<t_3<t_4$ gives rise to the 
integration measure
\begin{equation}
\int_0^\infty \dd t_2 
\int_{t_2}^\infty \dd t_3 
\int_{t_3}^\infty \dd t_4 
= \int_0^\infty \dd \Delta_1 
\int_{0}^\infty \dd \Delta_2 
\int_{0}^\infty \dd \Delta_3 \,.
\end{equation}
The final result is 
\begin{multline}
\label{finres4}
\int_{-\infty}^{+\infty} \dd t_2
\int_{-\infty}^{+\infty} \dd t_3
\int_{-\infty}^{+\infty} \dd t_4  \,  
C^{(4)}(t_2,t_3,t_4) \\
= 24 \sum_{n,n',n''>0}
\frac{\langle 0 |q|n \rangle \,
\langle n |q|n' \rangle \,
\langle n' |q|n'' \rangle \,
\langle n''|q|0 \rangle }{(E_n-E_0) \, (E_{n'}-E_0) \,
(E_{n''}-E_0) }
\\
- 24  \sum_{n,n'>0}
\frac{|\langle 0 |q|n \rangle|^2|\langle 0 |q|n' \rangle|^2 \, (E_{n}+E_{n'}-2E_0)}
{2(E_{n'}-E_0)^2(E_{n}-E_0)^2}
\\
= 24 \sum_{n,n',n''>0}
\frac{\langle 0 |q|n \rangle \,
\langle n |q|n' \rangle \,
\langle n' |q|n'' \rangle \,
\langle n''|q|0 \rangle }{(E_n-E_0) \, (E_{n'}-E_0) \,
(E_{n''}-E_0) }
\\
- 24  \sum_{n,n'>0}
\frac{|\langle 0 |q|n \rangle|^2|\langle 0 |q|n' \rangle|^2}%
{(E_{n'}-E_0)^2 \, (E_{n}-E_0)} \,.
\end{multline}
A remark is in order, here. 
The sums over intermediate states
in Eq.~\eqref{finres4}
exclude the ground state (which has $n=0$). 
One might wonder, for example, why the presence of the term
$- C^{(2)}(\Delta_1) \, C^{(2)}(\Delta_3)$
in Eq.~\eqref{connect4} does not lead to 
divergences, when the connected four-point
function is integrated over $\Delta_2$.
The term cancels, though, against the term
derived from the expression
$\langle 0 | q\, \ee^{-(H-E_0)\Delta_3}\,
q \, \ee^{-(H-E_0)\Delta_2}\, 
q \, \ee^{-(H-E_0)\Delta_1} q |0\rangle$
upon insertion of the ground state 
as the intermediate state in the 
exponential $\ee^{-(H-E_0)\Delta_2}$.
The term with the virtual ground state is thus excluded from
the sums over intermediate 
states in the representation~\eqref{finres4}.

Results for the 
perturbative coefficients
obtained from the four-point integral
\begin{multline} 
\int_{-\infty}^{+\infty} \dd t_2
\int_{-\infty}^{+\infty} \dd t_3
\int_{-\infty}^{+\infty} \dd t_4  \,
C^{(4)}(t_2,t_3,t_4) \\
\sim \sum_{K=0}^\infty [ G^{(4)}(1,1) ]_K \, g^K 
\end{multline} 
are summarized in Table~\ref{table1}.

The last correlation function 
studied here concerns the two-point function with a wigglet insertion
(see Ref.~\cite{GiEtAl2020}).
We have
\begin{multline}
C^{(1,2)}(t_2,t_3) =
\langle q(0)q(t_2)q^2(t_3)\rangle 
\\
- \langle q(0)q(t_2)\rangle \langle q^2(t_3)\rangle 
- 2\langle q(0)q(t_3)\rangle \langle q(t_2)q(t_3))\rangle \,,
\end{multline}
with the integral finding the perturbative expansion
(for some concrete sample values, see Table~\ref{table1})
\begin{equation}
\int_{-\infty}^{+\infty} \dd t_2
\int_{-\infty}^{+\infty} \dd t_3 \,  C^{(1,2)}(t_2,t_3) 
\sim \sum_{K=0}^\infty [ G^{(1,2)}(1,1) ]_K \, g^K \,.
\end{equation}
We obtain in terms of the perturbed oscillator
eigenstates,
\begin{multline}
\int_{-\infty}^{+\infty} \dd t_2
\int_{-\infty}^{+\infty} \dd t_3 \,  C^{(1,2)}(t_2,t_3) 
\\
= 4 \sum_{n,n'>0}
\frac{ \langle 0 |q^2|n \rangle \, 
\langle n |q|n' \rangle \, 
\langle n' |q|0 \rangle}{(E_n-E_0) \, (E_{n'}-E_0)}
\\
+ 2 \sum_{n,n'>0} \frac{\langle 0 |q|n \rangle
\langle n |q^2|n' \rangle
\langle n' |q|0 \rangle}{(E_n-E_0) \, (E_{n'}-E_0)}
\nonumber
\\
- 2 \sum_{n>0}
\frac{|\langle 0 |q|n \rangle|^2 \, \langle 0 |q^2|0 \rangle}{(E_n-E_0)^2}
\nonumber
\\
- 8  \sum_{n,n'>0}
\frac{|\langle 0 |q|n \rangle|^2|\langle 0 |q|n' \rangle|^2}%
{(E_{n'}-E_0) \, (E_{n'}+E_{n}-2E_0)}
\nonumber
\\
- 4 \left( \sum_{n>0} \frac{|\langle 0|q|n\rangle|^2}{E_n-E_0}  \right)^2 \, .
\end{multline}
The structures encountered for 
higher-dimensional internal symmetry groups 
are more complicated and will be discussed
in the following.
\color{black}

\begin{table*}[t!]
\begin{center}
\begin{minipage}{12cm}
\begin{center}
\caption{\label{table1} Sample values are 
collected for low-order perturbative coefficients 
of correlation function for the scalar theory ($N=1$).
Results are given as rational numbers when expressible
in compact form.}
\begin{tabular}{c@{\hspace{0.5cm}}c@{\hspace{0.5cm}}c@{\hspace{0.5cm}}%
c@{\hspace{0.5cm}}c}
\hline
\hline
\stdrule
$K$ & $[G^{(2)}(N,1)]_K$ & $[G^{(\partial p)}(N,1)]_K$ & %
$[G^{(4)}(N,1)]_K$ & $[G^{(2,1)}(N,1)]_K$ \\
\hline
\hline
\stdrule
0 & 1 & 2 & 0 & 0 \\
\stdrule
1 & $-\frac{3}{2}$ & $-6$ & $-6$  & $-\frac{3}{2}$ \\
\stdrule
2 & $\frac{31}{8}$ & $\frac{181}{9}$ & $\frac{99}{2}$ & $11$ \\
\stdrule
3 & $-\frac{1683}{128}$ & $-\frac{7397}{96}$ & $-\frac{663}{2}$ &
$-\frac{8805}{128}$ \\
\stdrule
4 & $\frac{13825}{256}$ & $\frac{9641561}{28800}$ & $\frac{277245}{128}$ & $\frac{55183}{128}$ \\
\stdrule
5 & $-258.264$ & $-1635.174$ & $-\frac{466581}{32}$ & $-2829.612$ \\
\stdrule
$\dots$ & & & & \\
\stdrule
10 & $3.540 \cdot 10^6$ & $2.115 \cdot 10^7$ & $5.003 \cdot 10^8$ & $9.473\cdot
10^7$ \\
\hline
\hline
\end{tabular}
\end{center}
\end{minipage}
\end{center}
\end{table*}

%
%
\section{Two-- and Three--Dimensional Quantum Anharmonic Oscillator}
\label{sec3}

%
%
\subsection{Overview}
\label{sec3A}

When $N>1$, since both the unperturbed Hamiltonian and 
the perturbation term are radially symmetric, we can use 
hyper-spherical coordinates
in order to describe the internal $O(N)$ space of the theory. 
The Hamiltonian is therefore
\begin{equation}
H = - \frac{1}{2} 
\left( \frac{\partial^2}{\partial r^2} 
+ \frac{N-1}{r} \frac{\partial}{\partial r} 
- \frac{\vec{L}^2}{r^2} \right)
+ \frac{r^2}{2} + \frac{g}{4} r^4,
\end{equation}
where $\vec{L}^2$ is the angular momentum in $N$ dimensions. The eigenfunctions
of the Hamiltonian can be written 
in terms of radial and angular parts, as follows,
\begin{equation}
\psi(\vec{r}) = R_{n \ell}(r) \, 
Y_\ell^{\ell_1 \dots , \ell_{N-2}}(\theta_1, \dots, \theta_{N-1}),
\end{equation}
where $Y_\ell^{\ell_1 \dots , \ell_{N-2}}(\theta_1, \dots, \theta_{N-1})$ are the
generalization of the spherical harmonics in $N$ dimensions and the
eigenfunctions of the angular momentum operator
\begin{multline}
\vec{L}^2 \, Y_\ell^{\ell_1 \dots , 
\ell_{N-2}}(\theta_1, \dots, \theta_{N-1}) \\
= \ell (\ell + N - 2) \, Y_\ell^{\ell_1 \dots , 
\ell_{N-2}}(\theta_1, \dots, \theta_{N-1}),
\end{multline}
with the quantum numbers $\ell_1 \dots , \ell_{N-2}$ satisfying 
$\left| \ell_{1}\right| \le |\ell_2| \le \dots \le | \ell_{N-2}| \le \ell$. 
Here, the angles $\theta_1$, $\dots$, $\theta_{N-2}$ range over
$[0,\pi]$, whereas $\theta_{N-1}$ ranges over
$[0,2\pi)$ (for a more thorough discussion, see Ref.~\cite{JeSa2018}).
In $N = 2$ dimensions, the unperturbed normalized radial functions 
$R^{(0)}_{n \ell}(r)$ 
\color{black}
and $Y_\ell^{\ell_1 \dots , \ell_{N-2}}(\theta_1, \dots, \theta_{N-1})$ 
are defined as~\cite{KaEtAl2014}
\begin{align}
Y_\ell(\theta_1) =& \; \frac{1}{\sqrt{2 \pi}} \, \ee^{\ii \ell \theta_1}, 
\\
R^{(0)}_{n \ell}(r) =& \; 
\calN^{(N=2)}_{n \ell} \, 
\exp\left( - \frac{r^2}{2} \right) \,
r^{|\ell|} \,
L^{|\ell|}_{\frac{1}{2}(n-|\ell|)}(r^2),
\label{eigen2d}
\end{align}
while in three dimensions, the expressions are 
as follows
($\ell_1 = m$ takes the role of the magnetic projection),
\begin{subequations}
\begin{align}
Y_\ell^{m}(\theta_1, \theta_2) = & \;
Y_{\ell m}(\theta = \theta_1, \varphi = \theta_2) \\[0.1133ex]
= & \; 
(-1)^m\sqrt{\frac{2 \ell+1}{4\pi}\frac{(\ell-m)!}{(\ell+m)!}} \,
\ee^{\ii m \theta_2} \, 
P_l^m(\cos \theta_1),
\nonumber\\
R^{(0)}_{n \ell}(r) =& \; \calN^{(N=3)}_{n \ell} \, r^{\ell} \,
\exp\left( - \frac{r^2}{2} \right) \,
L^{(\ell+\frac{1}{2})}_{\frac{1}{2}(n-\ell)}(r^2) \,.
\end{align}
\end{subequations}
Here, the $L_{k}^{{(\alpha )}}(r^{2})$ are the 
generalized Laguerre 
polynomials of order $k$, and the 
$P_\ell^m$ are the associated Legendre polynomials.
For three dimensions ($N = 3$), we indicate the relation 
of the $Y_\ell^{m}(\theta_1, \theta_2)$ to the 
more common notation 
$Y_{\ell m}(\theta = \theta_1, \varphi = \theta_2)$
of the spherical harmonics~\cite{VaMoKh1988},
where $\theta$ is the polar angle, 
and $\varphi$ is the azimuth angle.
The normalization factors are obtained as follows,
\color{black}
\begin{equation}
\begin{split}
& \calN^{(N=2)}_{n \ell} = 
\frac{\sqrt{2[(n-|\ell|)/2]!}}{\sqrt{[(n+|\ell|)/2]!}},\\
& \calN^{(N=3)}_{n \ell} = 
\left( \frac{2^{n + \ell + 2}}{\sqrt{\pi}}\right)^{1/2} 
\left[\frac{\Gamma\left(\frac{n-\ell}{2}+1\right) 
\Gamma\left(\frac{n+\ell}{2}+1\right)}%
{\Gamma\left(n+\ell+2\right)}\right]^{1/2}.
\label{eq::norm3N}
\end{split}
\end{equation}
Note that the generalized Laguerre polynomial satisfy the following 
orthogonality relation \cite{AbSt1972}
\begin{equation}
\int _{0}^{\infty } x^{\alpha } \, \ee^{-x}L_{n}^{(\alpha )}(x) \,
L_{m}^{(\alpha )}(x) \, \dd x = {\frac {(n+\alpha )!}{n!}} \, \delta _{n\,m} \,.
\end{equation}
It follows that the eigenfunctions are orthogonal,
\begin{equation}
\int\limits_0^\infty dr \, r^{N-1} 
R^{(0)}_{n \ell}(r) R^{(0)}_{s\ell}(r) = 
\frac{\left(\calN_{n \ell}^{(N)}\right)^2 \,
\Gamma\left(\frac{n+\ell+N}{2} \right)}%
{ 2 \, \Gamma\left(\frac{n-\ell}{2}+1 \right)} \,
\delta_{n \,s} \,,
\end{equation}
with $N=2,3$.
The perturbed Schr{\"o}dinger equation reduces 
to an equation for the radial part which reads
\begin{multline}
\label{eq:RadialEquation}
\left[ \frac{\dd^2}{\dd r^2} + 
\frac{N-1}{r} \frac{\dd}{\dd r} - 
\frac{\ell (\ell + N-2)}{r^2} + 
r^2 + \frac{g}{2} r^4\right] R_{nl}(r) 
\\
= -2 E_{n} R_{nl}(r).
\end{multline}
The matrix elements of the perturbation 
$\delta H = g \, r^4/4$, can be
written in the unperturbed basis evaluating the following integral
\begin{equation}
\bra{R^{(0)}_{n \ell}} r^4 \ket{R^{(0)}_{sr}} = 
\delta_{\ell \, r} \int_0^\infty \dd r \, r^{N+ 3}  \, 
R_{n \ell}^{(0)}(r) R_{s \ell}^{(0)}(r) \,.
\end{equation}
The eigenvalue relation given in Eq.~\eqref{eq:RadialEquation} 
can be written in the space of radial functions only,
\begin{equation}
(H_0+\delta H)(R^{(0)}_{n \ell} +
\delta R_{n \ell})=(E^{(0)}_{n}+\delta E_{n \ell}) \,
(R^{(0)}_{n \ell} + \delta R_{n \ell}).
\label{Shrodinger_pert}
\end{equation}
The perturbation does not change the angular part, but only the radial
one. So, if we know the unperturbed radial part $R^{(0)}_{n \ell}(r)$,
then we can compute the
perturbation to the energy $\delta E_{n \ell}$ and to the eigenfunction 
$\delta R_{n \ell}$, for fixed value of $\ell$. 
In fact, for fixed $\ell$,
the spectrum is not degenerate, and we can apply standard perturbation
theory, as we did in the previous section for the case $N=1$. 

A number of useful formulas for the treatment of the 
$O(N)$ problem, both in terms of the angular 
algebra as well as the perturbative treatment of the 
radial part, are given in Appendices~\ref{app:identities}
and~\ref{alternative}.
\color{black}

\begin{table*}[t!]
\caption{\label{table2} Same as Table~\ref{table1}, but for the 
$N=2$ theory.}
\begin{tabular}{c@{\hspace{0.5cm}}c@{\hspace{0.5cm}}c@{\hspace{0.5cm}}%
c@{\hspace{0.5cm}}c}
\hline
\hline
\stdrule
$K$ & $[G^{(2)}(N,1)]_K$ & $[G^{(\partial p)}(N,1)]_K$ & %
$[G^{(4)}(N,1)]_K$ & $[G^{(2,1)}(N,1)]_K$ \\
\hline
\hline
\stdrule
0 & 1 & 2 & 0 & 0 \\
\stdrule
1 & $-2$ & $-8$ & $-6$  & $-\frac{3}{2}$ \\
\stdrule
2 & $\frac{20}{3}$ & $\frac{940}{27}$ & $63$ & $\frac{503}{36}$ \\
\stdrule
3 & $-\frac{2041}{72}$ & $-\frac{9113}{54}$ & $-\frac{3151}{6}$ &
$-108.435$ \\
\stdrule
4 & 142.078 & 904.550 & $4179.733$ & $824.861$ \\
\stdrule
5 & $-809.263$ & $-5318.618$ & $-33562.531$ & $-6433.962$ \\
\stdrule
$\dots$ & & & & \\
\stdrule
10 & $1.993 \cdot 10^7$ & $1.267 \cdot 10^8$ & $2.114 \cdot 10^9$ & $3.906\cdot
10^8$ \\
\hline
\hline
\end{tabular}
\end{table*}

From the knowledge of the eigenfunctions and eigenvalues of the Hamiltonian,
we can determine the $M$-point correlation functions of our theory given by
\cite{ZJ2021}
\begin{multline}
C_{i_1 \, i_2 \, ... \, i_{M}}(t_1,t_2,...,t_{M}) \\
\equiv \langle \phi_{i_1}(t_1)\,\phi_{i_2}(t_2)\,\cdots\,\phi_{i_M}(t_{M}) \rangle_C\\
\equiv T_{i_1 \, i_2 \,\dots \,i_M} \, 
C^{(M)}(t_2 - t_1,...,t_{M} - t_1),
\label{corrM1}
\end{multline}
where in the latter form, the expression
$C^{(M)}(t_2 - t_1,...,t_{M} - t_1)$
has $M-1$ arguments and is defined in 
analogy to the four-point correlation 
function from Eq.~\eqref{connect4}.
For example, we have $M = 2$ for the two-point function,
and one argument in $C^{(2)}(t_2 - t_1)$,
while, of course, we have $M = 4$ for the four-point function,
and three arguments in 
$C^{(4)}(t_2 - t_1, t_3 - t_1, t_4 - t_1)$.
Furthermore, the indices $i_j$ with $j = 1, \ldots, M$ 
can obtain values $1 \leq i_j \leq N$,
consistent with the structure of the 
internal symmetry group.
The designation $C$ indicates that we are 
considering the connected part of correlation function, and
$T_{i_1, i_2,\dots,i_{M}}$ is the average value of the product of 
unit vector components
$u_{i_1}$, $u_{i_2}$, and so on, taken over the 
the $N$-dimensional unit sphere,
\begin{equation}
T_{i_1 \, i_2 \,\dots \,i_{M}} =
\langle u_{i_1} \, u_{i_2} \,\dots \,u_{i_M}\rangle_{S_{N-1}} \,.
\end{equation}
with $S_{N-1}$ being the $N-1$-dimensional 
unit sphere embedded in $N$ dimensions;
so,  $S_{N-1}$ has
a (generalized) surface volume $\Omega_N = 2 \pi^{N/2}/\Gamma(N/2)$. 
We can write an $M$-point correlation function with arbitrary indices
$i_1, i_2,\dots,i_{M}$ in terms of the same correlation function with fixed
indices $\hat{\imath}_1,\hat{\imath}_2,\dots,\hat{\imath}_M$ by multiplying and
dividing the previous expression by
$T_{\hat{\imath}_1,\hat{\imath}_2,\dots,\hat{\imath}_M}$ 
\begin{multline}
C_{i_1 \, i_2 \,\dots \,i_{M}}(t_1,t_2,\dots,t_{M}) \\
\equiv \frac{T_{i_1 \, i_2\,\dots \,i_{M}}}%
{T_{\hat{\imath}_1 \, \hat{\imath}_2 \, \dots,\hat{\imath}_M}} \;
C_{\hat{\imath}_1 \,\hat{\imath}_2 \,\dots \, \hat{\imath}_M}(t_1,t_2,\dots,t_{M}) \,.
\label{corrM2}
\end{multline}
Here, we write the indices of the fixed element
$C_{\hat{\imath}_1,\hat{\imath}_2,\dots,\hat{\imath}_M}(t_1,t_2,\dots,t_{M})$
with a hat. Our convention is that 
indices with hats are not being summed over,
even when they are repeated (in other words, we use the 
convention that the Einstein summation convention
does not apply on indices with hats).
\color{black}
Comparing Eq.~\eqref{corrM1} to Eq.~\eqref{corrM2}, we can write
\begin{equation}
C^{(M)}(t_2 - t_1,...,t_{M} - t_1) =
\frac{C_{\hat{\imath}_1 \,\hat{\imath}_2 \,\dots 
\,\hat{\imath}_M}(t_1,t_2,\dots,t_{M})}%
{T_{\hat{\imath}_1,\hat{\imath}_2,\dots,\hat{\imath}_M}} \,.
\end{equation}
This formula allows us to pick one single nonvanishing
element of the correlation function,
say, one where all indices are equal,
\begin{equation}
\hat{\imath}_1 = \hat{\imath}_2 = \dots \hat{\imath}_M \,,
\end{equation}
and to derive a valid expression for any 
combination of the $\hat{\imath}_j$, 
with $j = 1, \dots, M$. 
It is useful to recall the well-known results~\cite{ZJ2021}
\begin{equation}
T_{i_1 \, i_2} 
= \langle u_{i_1}\, u_{i_2} \rangle_{S_{N-1}} 
= \frac{\delta_{i_1 \, i_2}}{N} \,,
\end{equation}
and
\begin{multline}
T_{i_1 \, i_2 \, i_3 \, i_4} 
= \langle u_{i_1}\, u_{i_2} \, u_{i_3} \, u_{i_4} \rangle_{S_{N-1}} 
\\
= \frac{\delta_{i_1 \, i_2} \delta_{i_3 \, i_4} +
\delta_{i_1 \, i_3} \delta_{i_2 \, i_4} +
\delta_{i_1 \, i_4} \delta_{i_2 \, i_3} }{N \, (N + 2)} \,.
\end{multline}
In this way, the four-point function written only in terms 
of the element with all indices fixed to $\hat{\imath}_1$,
becomes
\begin{multline}
C_{i_1, i_2, i_3, i_4}( t_1, t_2, t_3, t_4 ) \\ 
\equiv \frac{\delta_{i_1\,i_2}\delta_{i_3\,i_4} +
\delta_{i_1\,i_3}\delta_{i_2,i_4} +
\delta_{i_1\,i_4}\delta_{i_2\,i_3}}{3} \\
\times C_{\hat{\imath}_1 \, \hat{\imath}_1 \, \hat{\imath}_1 \, 
\hat{\imath}_1}(t_1,t_2,t_3,t_{4})\\ 
\equiv \frac{\delta_{i_1 \,i_2} \delta_{i_3 \, i_4} +
\delta_{i_1 \, i_3}\delta_{i_2 \, i_4} +
\delta_{i_1 \, i_4}\delta_{i_2 \, i_3}}{N(N+2)}\\ 
\times \left[\frac{N(N+2)}{3}
C_{\hat{\imath}_1 \, \hat{\imath}_1 \, 
\hat{\imath}_1 \, \hat{\imath}_1}(t_1,t_2,t_3,t_4) \right] \\
\equiv \frac{\delta_{i_1 \,i_2} \delta_{i_3 \, i_4} +
\delta_{i_1 \, i_3}\delta_{i_2 \, i_4} +
\delta_{i_1 \, i_4}\delta_{i_2 \, i_3}}{N(N+2)}\\
\times C^{(4)}(t_2 - t_1,t_3 - t_1,t_4 - t_1) \,.
\end{multline}
So, we have
\begin{multline}
\label{angular_reduction}
C^{(4)}(t_2 - t_1,t_3 - t_1,t_4 - t_1) = \frac{N(N+2)}{3} 
\\
\times 
C_{\hat{\imath}_1 \, \hat{\imath}_1 \, \hat{\imath}_1 \, \hat{\imath}_1}%
(t_1,t_2,t_3,t_4) 
\\
\equiv
\frac{N(N+2)}{3} 
\widetilde{C}^{(4)}(t_2 - t_1,t_3 -t_1,t_4 - t_1) \,,
\end{multline}
where the latter expression defines the 
quantity $\widetilde{C}^{(4)}(t_2 - t_1,t_3 -t_1,t_4 - t_1) =
C_{\hat{\imath}_1 \, \hat{\imath}_1 \, \hat{\imath}_1 \, \hat{\imath}_1}%
(t_1,t_2,t_3,t_4)$.

\color{black}

%
%
\subsection{Two-point correlator and second derivative}
\label{sec3B}

We start the discussion from the two-point correlation function and its second
derivative with respect to the momenta. Changing coordinates and picking up one
of the possible components (because they all give the same contribution) we get
\begin{equation}
C_{i_1 \,i_2}(t) = \frac{\delta_{i_1\,i_2}}{N} \, C^{(2)}(t),  \\
\end{equation}
with
\begin{equation}
\label{two_points}
C^{(2)}(t)\equiv N\langle q_{\hat{\imath}_1}(0) 
q_{\hat{\imath}_1}(t)\rangle 
= N \,  C_{\hat{\imath}_1\, \hat{\imath}_1}(t)
= N \, \widetilde{C}^{(2)}(t)\,,
\\[0.1133ex]
\end{equation}
where $\widetilde{C}^{(2)}(t) = 
C^{(2)}_{\hat{\imath}_1\, \hat{\imath}_1}(t)$
is equal to any nonvanishing element within the 
internal group structure.
\color{black}
So, $\hat{\imath}_1$ can assume one of the $N$ possible values. We can
therefore choose for example $\hat{\imath}_1 = 1$, and write
\begin{multline}
\label{C2angular}
\langle q_{\hat{\imath}_1=1}(0) q_{\hat{\imath}_1=1}(t)\rangle
= \langle 0 |r \cos \theta_1 \, 
\ee^{-(H-E_0) |t|} \, r \cos \theta_1 |0\rangle 
\\[0.1133ex]
= \sum_{\vec{\eta}} |\langle 0|r \cos \theta_1| 
n, \vec{\eta} \rangle|^2 \,
\ee^{-(E_{n}-E_0) |t|} \,.
\end{multline}
We have defined $| \vec{\eta}\rangle =
|\ell, \ell_1, \dots, \ell_{N-2}\rangle$
as a vector representing all the angular quantum 
numbers of the state.
The state $| n, \vec \eta \rangle$ is 
characterized by the principal quantum number $n$,
and the set of angular quantum numbers $\vec \eta$.
For the cases $N=2$ and $N=3$ under investigation 
here, the angular integration lead to the 
following picture. First, one observes that 
the summation over all possible 
quantum number summarized in $\vec \eta$
selects those states which are 
coupled to the ground state by a 
dipole transition. We define
the coordinate system so that (in $N=3$) 
the quantization axis is aligned with the 
coordinate $i_1 = 1$, so that 
$r \, \cos\theta_1$ is the coordinate 
along the quantization axis.
In $N=2$, there exists no quantum number $\ell_1$;
we only have one angular momentum quantum 
number, $\ell$, which can take either positive or
negative integer values.
By contrast, for $N=3$, one has two quantum 
numbers, namely, the angular monentum
$\ell$ and the magnetic projection $\ell_1 = m$.

For $N=2$, there are two nonvanishing contributions 
to the sum over intermediate states in Eq.~\eqref{C2angular},
namely, those with $\ell = \pm 1$.  
The contributions from $\ell = -1$ and $\ell = +1$ 
are equal
to each other and can be taken into account
on the basis of an additional multiplicity factor.
For $N=3$, instead, the only nonvanishing contribution
to the sum over intermediate states in 
Eq.~\eqref{C2angular} is
given by states with $\ell=1$. 
States with $\ell = 1$ 
are commonly referred to as 
$P$ states in atomic physics~\cite{BeSa1957},
and these can have three angular momentum projections,
namely $\ell_1 = m = -1, 0, 1$.
Of theses, under an appropriate identification 
of the quantization axis, only the state with 
$\ell_1 = m = 0$ contributes, being coupled 
by the operator $r \, \cos\theta_1$. However, as is well 
known from atomic physics~\cite{BeSa1957}, the 
final result is independent of the choice of the 
quantization axis, provided one sums over $m$,
viz., sums over $\vec\eta$.

The two-point correlation function at zero momentum transfer
is obtained by integrating Eq.~\eqref{two_points} with respect to time,
\begin{equation}
\label{C2int}
\int_{-\infty}^{+\infty} \widetilde{C}^{(2)}(t) \, \dd t = 
2 \sum_{n, \vec{\eta}}
\frac{|\langle 0|r \cos \theta_1| n, \vec{\eta}\rangle|^2}{E_{n}-E_0} \,.
\end{equation}
Similarly, we get its second derivative at zero momentum transfer as
\begin{equation}
\label{C2derivint}
\int_{-\infty}^{+\infty} t^2\,
\widetilde{C}^{(2)}(t) \, \dd t = 
4 \sum_{n, \vec\eta}
\frac{|\langle 0|r \cos \theta_1|\vec{\eta}\rangle|^2}{(E_{n}-E_0)^3}.
\end{equation}
As discussed, each matrix element 
$\langle 0|r \cos(\theta)| n, \vec{\eta}\rangle$ 
can be computed using Eq.~\eqref{matrix_element1},
in terms of a radial transition 
matrix element $S_{00, n \ell}$,
and an angular element $\alpha^{(N)}_{0, \ell}$,
which depends on the dimension $N$. 
After the summation over the angular quantum 
numbers, one obtains the result
\begin{equation}
\begin{split}
\sum_{\vec\eta} 
| \langle 0|r \cos(\theta_1)| n, \vec{\eta}\rangle |^2 
= \widetilde{\sum_\ell} \left( S_{00, n \ell} \right)^2 \, 
\left( \alpha^{(N)}_{0, \ell} \right)^2 \,.
\label{0_cos_eta}
\end{split}
\end{equation}
where $\widetilde{\sum_\ell}$ is a sum 
over a single term $\ell = 1$ for $N=3$,
and over $\ell = \pm 1$ for $N=2$.

\begin{table*}[t!]
\caption{\label{table3} Same as Tables~\ref{table1} and~\ref{table2}, 
but for the $N=3$ theory.}
\begin{tabular}{c@{\hspace{0.5cm}}c@{\hspace{0.5cm}}c@{\hspace{0.5cm}}%
c@{\hspace{0.5cm}}c}
\hline
\hline
\stdrule
$K$ & $[G^{(2)}(N,1)]_K$ & $[G^{(\partial p)}(N,1)]_K$ & %
$[G^{(4)}(N,1)]_K$ & $[G^{(2,1)}(N,1)]_K$ \\
\hline
\hline
\stdrule
0 & 1 & 2 & 0 & 0 \\
\stdrule
1 & $-\frac{5}{2}$ & $-10$ & $-6$  & $-\frac{3}{2}$ \\
\stdrule
2 & $\frac{245}{24}$ & $\frac{1445}{27}$ & $\frac{153}{2}$ & $\frac{305}{18}$ \\
\stdrule
3 & $-\frac{60175}{1152}$ & $-\frac{271385}{864}$ & $-\frac{4583}{6}$ &
$-157.199$ \\
\stdrule
4 & 309.971 & 2007.796 & $7184.900$ & $1409.880$ \\
\stdrule
5 & $-2058.802$ & $-13870.761$ & $-67294.482$ & $-12793.097$ \\
\stdrule
$\dots$ & & & & \\
\stdrule
10 & $8.948 \cdot 10^7$ & $5.986 \cdot 10^8$ & $7.588 \cdot 10^9$ & $1.374 \cdot
10^9$ \\
\hline
\hline
\end{tabular}
\end{table*}

It is convenient to define the quantity
\begin{equation}
\label{alphatilde_def}
\widetilde{\alpha}^{(N)}_0 = 
\begin{cases}
4 \left( \alpha_{0,1}^{(2)}\right)^4 & \text{for } N=2\,, \\
\,\left( \alpha_{0,1}^{(3)}\right)^4 & \text{for } N=3 \,.
\end{cases}
\end{equation}
Calculating the square root of $\widetilde{\alpha}^{(N)}_0$,
the previously mentioned multiplicity factor two 
for $N=2$ is obtained; it takes care of the 
two equivalent contributions from $\ell = \pm 1$.
We therefore have for the two-point 
correlation function computed at zero momentum,
\begin{align}
\int_{-\infty}^{+\infty} \widetilde{C}^{(2)}(t) \, \dd t = & \;
2\, \sqrt{\widetilde \alpha_0^{(N)}} \sum_{n} {S_{00, n1}^2 \over E_{n}-E_0}
\nonumber\\[0.1133ex]
\sim & \;
\frac{1}{N} \, \sum_K [G^{(2)}(N,1)]_K \, g^K \,.
\end{align}
Perturbative coefficients $[G^{(2)}(N,1)]_K$ 
for $N=2$ and $N=3$ are
given in Tables~\ref{table2} and~\ref{table3}, 
respectively. 
The second derivative of the two-point correlator
is given as follows,
\begin{align}
\int_{-\infty}^{+\infty} t^2 \, 
\widetilde{C}^{(2)}(t) \, \dd t = & \;
4\, \sqrt{\widetilde \alpha_0^{(N)}} \sum_{n}  {S_{00,n1}^2 \over (E_{n}-E_0)^3} \,.
\nonumber\\[0.1133ex]
\sim & \;
\frac{1}{N} \, \sum_K [G^{(\partial p)}(N,1)]_K \, g^K \,.
\end{align}
Again, low-order perturbative coefficients 
$[G^{(\partial p)}(N,1)]_K$ for $N=2$ and $N=2$ are
given in Tables~\ref{table2} and~\ref{table3},
respectively.
At this point, all matrix elements have been reduced 
to radial integrals of
unperturbed $O(N)$ oscillator eigenstates.
In the evaluation of perturbative coefficients
of the energy levels of a number of anharmonic oscillators,
recursion relations have been 
found~\cite{ZJ1981jmp,ZJ1984jmp}.
For the quartic $O(N)$ oscillator, 
we refer to Eqs.~(22) and~(23) of Ref.~\cite{ZJ1981jmp},
for the double-well potential 
to Eqs.~(69) and~(70) of Ref.~\cite{ZJ1981jmp},
and for more general potentials, 
to Eqs.~(12)---(25) of Ref.~\cite{ZJ1984jmp}.
We do not attempt to generalize the treatment 
outlined in Refs.~\cite{ZJ1981jmp,ZJ1984jmp}
to the correlation functions investigated here, 
because the number of higher-order 
perturbative coefficients obtained 
using the methods delineated here is fully sufficient
for detecting the two-loop corrections to the 
large-order behavior of the perturbative coefficients~\cite{GiEtAl2020}.
For possible future investigations, we note that 
it would be interesting to explore 
recursion relations for the 
perturbative correlation functions.

\begin{figure}[t!]
\begin{center}
\includegraphics[width=0.9\linewidth]{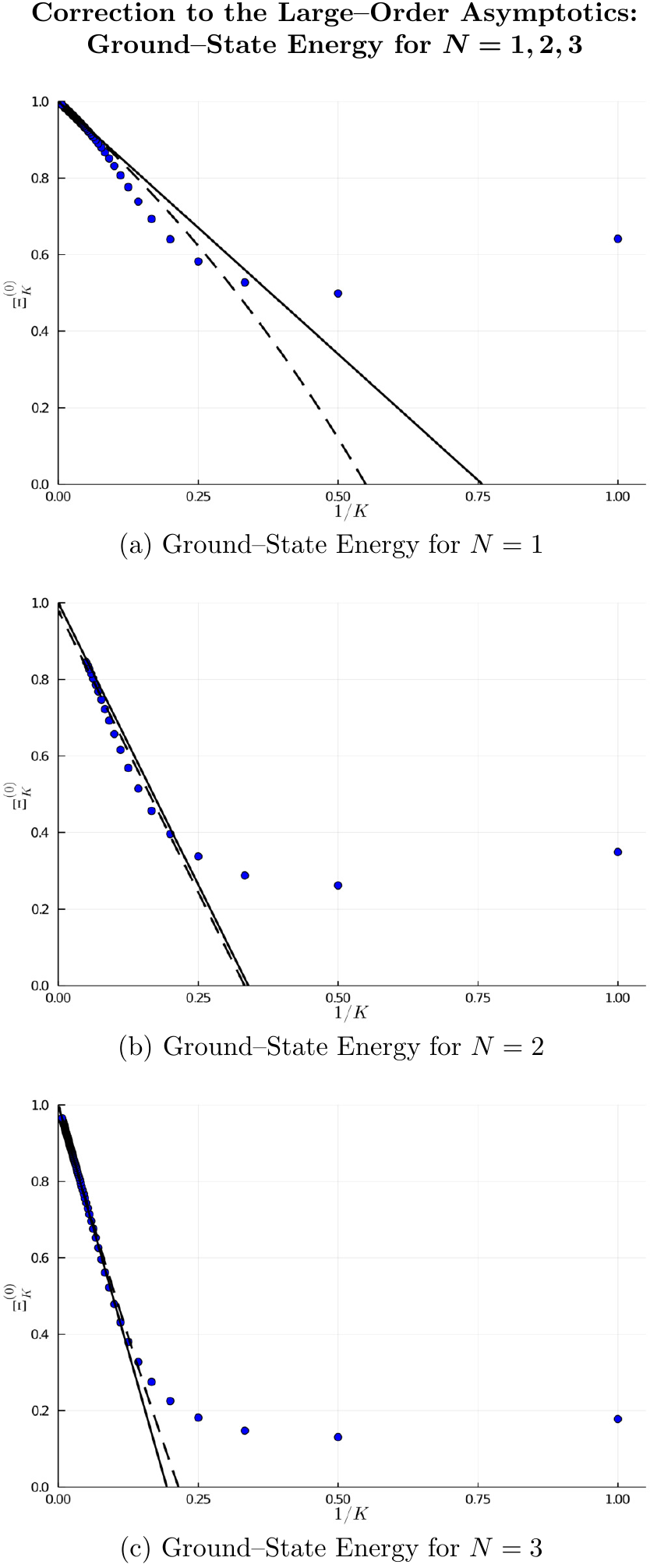}
\caption{\label{Fig1}
Coefficients of the perturbative expansion of the ground-state of the energy
for $N=1,N=2, N=3$ in function of the inverse of the order of perturbation
(blue dots). These coefficients have been divided by their leading asymptotic
estimate, given in Eq.~(\ref{XiM}), and have been compared with their 
subleading order estimate. 
Dashed black lines 
include the term $b-1$ term at the denominator
of the $1/(K + b - 1)$ term in Eq.~\eqref{nu}, 
while solid black lines exclude this term,
and follow only the $1/K$ term given in Eq.~\eqref{mu}.}
\end{center}
\end{figure}

%
%
\subsection{Four-point correlation function}
\label{sec3C}

Similarly to the previous subsection, 
we apply the same ideas to the 
connected four-point function, 
\begin{multline}
C_{i_1 \, i_2 \, i_3 \, i_4}(t_1,t_2,t_3,t_4) = \frac{1}{N(N+2)} \\
\times
( \delta_{i_1 \, i_2} \, \delta_{i_3\,i_4} +
\delta_{i_1 \, i_3} \, \delta_{i_2 \, i_4} +
\delta_{i_1 \, i_4} \, \delta_{i_2 \, i_3} ) \\ 
\times C^{(4)}(t_2 - t_1,t_3 - t_1,t_4 - t_1) \,,
\end{multline}
where, according to Eq.~\eqref{angular_reduction},
\color{black}
\begin{multline}
C^{(4)}(t_2,t_3,t_4) = 
\frac{N(N+2)}{3} \, \widetilde{C}^{(4)}(t_2,t_3,t_4) \\ 
\equiv \frac{N(N+2)}{3} \bigg[\langle q_{\hat{\imath}_1}(0) 
q_{\hat{\imath}_1}(t_2) q_{\hat{\imath}_1}(t_3) q_{\hat{\imath}_1}(t_4) \rangle 
\\
- \widetilde{C}^{(2)}(-t_2)\widetilde{C}^{(2)}(t_3-t_4) 
- \widetilde{C}^{(2)}(t_3-t_2)\widetilde{C}^{(2)}(-t_4) 
\\
- \widetilde{C}^{(2)}(-t_3)\widetilde{C}^{(2)}(t_2-t_4) \bigg]\,.
\end{multline}
In our derivation leading to Eq.~\eqref{angular_reduction},
we had stressed that the formula is valid for every 
value that $\hat{\imath}_1$ can assume. 
We now select, for convenience, 
one particular component with $\hat{\imath}_1=1$, 
which, as we assume, is aligned with the quantization axis
for the states in the internal $O(N)$ symmetry group. 
Hence, we write the relation
\color{black}
\begin{multline}
\langle q_{\hat{\imath}_1=1}(0) q_{\hat{\imath}_1=1}(t_2) 
q_{\hat{\imath}_1=1}(t_3) q_{\hat{\imath}_1=1}(t_4) \rangle
\\
= \langle 0 |r \cos \theta_1 \ee^{-(H-E_0)\Delta_3}\,
r \cos \theta_1\,
\ee^{-(H-E_0)\Delta_2} \\
r \cos \theta_1\, 
\ee^{-(H-E_0)\Delta_1} 
r \cos \theta_1 |0\rangle +
\nonumber
\\
-\widetilde{C}^{(2)}(\Delta_1)
\widetilde{C}^{(2)}(\Delta_3)-\widetilde{C}^{(2)}(\Delta_1+
\Delta_2)\widetilde{C}^{(2)}(\Delta_2+\Delta_3)
\\
- \widetilde{C}^{(2)}(\Delta_1+\Delta_2+\Delta_3)\widetilde{C}^{(2)}(\Delta_2),
\end{multline}
\color{black}
where $\Delta_i\equiv t_{i+1}-t_i$.
We then have, with $x_1 = r \cos \theta_1$,
\begin{multline}
\int_{-\infty}^{+\infty} \dd t_2
\int_{-\infty}^{+\infty} \dd t_3
\int_{-\infty}^{+\infty} \dd t_4 \,  \widetilde{C}^{(4)}(t_2,t_3,t_4)  =
\\
= 24 \sum_{\vec{\eta},\vec{\eta}',\vec{\eta''}} 
\frac{ \langle 0 | x_1 |\vec{\eta} \rangle
\langle \vec{\eta} | x_1 |\vec{\eta}'\rangle 
\langle \vec{\eta}' | x_1 |\vec{\eta''} \rangle
\langle \vec{\eta''}| x_1 |0 \rangle }
{(E_n-E_0) \, (E_{n'}-E_0) \, (E_{n''}-E_0)} 
\\
-24 \sum_{\vec{\eta},\vec{\eta}'}
\frac{ |\langle 0 | x_1|\vec{\eta} \rangle|^2 \,
|\langle 0 | x_1 |\vec{\eta}' \rangle|^2 (E_{n}+E_{n'}-2E_0)}{2(E_{n'}-E_0)^2(E_{n}-E_0)^2} \,.
\label{4point}
\end{multline}
Using Eq.~\eqref{matrix_element1}, we have
\begin{multline}
\int_{-\infty}^{+\infty} \dd t_2
\int_{-\infty}^{+\infty} \dd t_3
\int_{-\infty}^{+\infty} \dd t_4 \, 
\widetilde{C}^{(4)}(t_2,t_3,t_4)
\\ 
= 24 
\sum_{n, \ell, n', \ell'} 
\frac{1}{E_{n'}-E_0} 
\left[ \alpha^{(N)}_{0,\ell} \, 
\alpha^{(N)}_{\ell,\ell'} \frac{S_{00, n\ell} 
S_{n \ell, n'\ell'} }{E_n-E_0} \right]^2  \\
-24  
\sum_{n, \ell, n', \ell'}
\frac{ \left(\alpha^{(N)}_{0,\ell'}\right)^2 S^2_{00, n\ell} \, 
S^2_{00, n'\ell'}}{ (E_{n'}-E_0)^2 \, (E_{n}-E_0)} \,.
\color{black}
\end{multline}

The multiplicities are different if we choose $N=2$ or $N=3$. 
We can insert them in the definition of the quantity
\begin{equation}
\label{alphatilde_redef}
\widetilde{\alpha}^{(N)}_\ell = 
\begin{cases}
4 \left( \alpha_{0,1}^{(2)}\right)^4 \delta_{\ell, 0} + 
2 \left( \alpha_{0,1}^{(2)} \alpha_{1,2}^{(2)} \right)^2 \delta_{\ell, 2} & 
\text{for } N=2\,, \\
\left( \alpha_{0,1}^{(3)}\right)^4 \delta_{\ell, 0} + 
\left( \alpha_{0,1}^{(3)} \alpha_{1,2}^{(3)} \right)^2 \delta_{\ell, 2} & \text{for } N=3 \,,
\end{cases}
\end{equation}
which depend on the factors given in Eqs.~\eqref{eq::alpha0}, \eqref{eq::alpha1}. 
Using the angular selection rules,
we finally obtain
\begin{multline}
\label{C4int}
\int_{-\infty}^{+\infty} \dd t_2
\int_{-\infty}^{+\infty} \dd t_3
\int_{-\infty}^{+\infty} \dd t_4 \, 
\widetilde{C}^{(4)}(t_2,t_3,t_4)
\\
= 24 \sum_{n'} \sum_{\ell'=0, 2} \frac{\widetilde \alpha^{(N)}_{\ell'}}{E_{n'}-E_0} 
\left[ \sum_{n} \frac{S_{00, n1} S_{n1, n' \ell'} }{E_n-E_0} \right]^2  
\\
-24 \, \widetilde \alpha^{(N)}_{0} \sum_{n, n'} \frac{ S^2_{00, n1} \, 
S^2_{00, n'1} (E_{n}+E_{n'}-2E_0)}{ 2(E_{n'}-E_0)^2(E_{n}-E_0)^2} 
\\
\sim \frac{3}{N(N+2)} \, \sum_K [ G^{(4)}(N,1)]_K \, g^K \,.
\end{multline}
Results for low-order perturbative coefficients
for the four-point correlation function
($N=2$ and $N=3$)
are given in Tables~\ref{table2} and~\ref{table3}.
\color{black}

%
%
\subsection{Correlation function with a wigglet insertion}
\label{sec3D}

The last quantity we want to look at,
within the context of the $O(N)$ theory, 
is the two-point correlation 
function with a wigglet insertion,
\begin{equation}
C^{(1,2)}_{i_1,i_2}(t_2,t_3) =
\frac{\delta_{i_1 \,i_2}}{N} \, C^{(1,2)}(t_2, t_3) \,,
\label{eq:wigglet1}
\end{equation}
where we have defined
\begin{equation}
\label{wigglet_ON}
C^{(1,2)}(t_2, t_3) \equiv 
N \,  C^{(1,2)}_{\hat{\imath}_1\, \hat{\imath}_1}(t_2, t_3)
= N \, \widetilde{C}^{(1,2)}(t_2, t_3)\,.
\\[0.1133ex]
\end{equation}
Here, $\widetilde{C}^{(1,2)}(t_2, t_3) = 
C^{(1,2)}_{\hat{\imath}_1\, \hat{\imath}_1}(t_2, t_3)$
is equal to any nonvanishing element within the
internal group structure;
nonvanishing elements have their first index
equal to their second index.
One finds
\begin{multline}
\widetilde{C}^{(1,2)}(t_2,t_3) =
\langle q_{\hat{\imath}_1}(0) \, q_{\hat{\imath}_1}(t_2) \,
q_{\hat{\imath}_1}^2(t_3)\rangle
\\
-\langle q_{\hat{\imath}_1}(0) \, q_{\hat{\imath}_1}(t_2)\rangle 
\, \langle q_{\hat{\imath}_1}^2(t_3)\rangle
\\
-2\langle q_{\hat{\imath}_1}(0) \, q_{\hat{\imath}_1}(t_3)\rangle 
\langle q_{\hat{\imath}_1}(t_2) \, q_{\hat{\imath}_1}(t_3))\rangle \,.
\end{multline}
Setting $\hat{\imath}_1 = 1$ for simplicity and introducing the 
quantity $x_1 = r \, \cos\theta_1$, for the coordinate along the 
quantization axis, as before, we can perform the integrals 
with respect to the time variables, obtaining
\begin{multline} 
\int_{-\infty}^{+\infty} \dd t_2\int_{-\infty}^{+\infty} \dd t_3 \, 
\widetilde{C}^{(1,2)}(t_2,t_3) \\
= 4 \sum_{n,\vec{\eta},n',\vec{\eta}'} 
\frac{ \langle 0 | (x_1)^2 | n, \vec{\eta} \rangle
\langle n, \vec{\eta} | x_1| n', \vec{\eta}' \rangle
\langle n', \vec{\eta}' | x_1| n, 0 \rangle }%
{ ( E_n-E_0 ) \, ( E_{n'}-E_0 ) }
\\ 
+ 2 \sum_{n,\vec{\eta},n',\vec{\eta}'}
\frac{\langle 0 | x_1 | n, \vec{\eta} \rangle
\langle n, \vec{\eta} | (x_1)^2 | n', \vec{\eta}' \rangle
\langle n', \vec{\eta}' | x_1 | 0 \rangle }
{ (E_n-E_0) \, (E_{n'}-E_0) }
+
\\
- 2 \sum_{n, \vec{\eta}}
\frac{ |\langle 0 | \, x_1 \, | n, \vec{\eta} \rangle|^2
\langle 0 | \, (x_1)^2 \, |0 \rangle  }{(E_n-E_0)^2}
- 4 \left( \sum_{ n, \vec{\eta} }
\frac{ |\langle 0| \, x_1 \, | n, \vec{\eta} \rangle|^2}{(E_n-E_0)}  \right)^2 
\\
- 8 \sum_{ n, \vec{\eta}, n', \vec{\eta}'}
\frac{ |\langle 0 | x_1 | n, \vec{\eta} \rangle|^2| \,
\langle 0 | r \cos \theta_1| n', \vec{\eta}' \rangle|^2 }%
{ (E_{n'}-E_0) \, (E_{n'}+E_{n}-2E_0)} \,.
\label{wigg1}
\end{multline} 
Using the radial $S$ matrix
elements defined in Eq.~\eqref{matrix_element1}, 
and the radial $Q$ matrix elements
defined in Eq.~\eqref{matrix_element2},
we obtain
\begin{multline}
\int_{-\infty}^{+\infty} \dd t_2
\int_{-\infty}^{+\infty} \dd t_3 \,  \widetilde{C}^{(1,2)}(t_2,t_3) =
\\
4 (2-\delta_{N\,3})\sum_{n, n'} \sum_{\ell=0,2} 
\frac{ \beta_{0,\ell}^{(N)} 
\alpha_{\ell,1}^{(N)} \alpha_{1,0}^{(N)} \, Q_{00, n\ell} S_{n\ell,n'1} 
S_{n'1,00}}{(E_n-E_0)(E_{n'}-E_{0})} \\
+ 2 (3-2\delta_{N\,3})\sum_{n, n'} 
\frac{ \beta_{1,1}^{(N)} 
\left(\alpha_{1,0}^{(N)}\right)^2 \, S_{00, n1} 
Q_{n1,n'1} S_{n'1,00}}{(E_n-E_0)(E_{n'}-E_{0})} 
\\
- 2 (2-\delta_{N\,3})\sum_{n} \left(\alpha_{0,1}^{(N)}\right)^2 
\beta_{0,0}^{(N)} \frac{S^2_{00, n1} Q_{00,00}}{(E_{n}-E_0)^2}\\
- 8  (2-\delta_{N\,3})^2\sum_{n, n'} 
\frac{ \left(\alpha_{0,1}^{(N)}\right)^2 \,
\left(\alpha_{0,1}^{(N)}\right)^2  \, S^2_{00, n1} S^{2}_{00,n'1}}
{(E_{n'}-E_0)(E_{n'}+E_{n}-2E_{0})}\\
-4 (2-\delta_{N\,3})^2
\left[ \sum_{n} \left(\alpha_{0,1}^{(N)}\right)^2 \frac{S_{00, n1}^2}{E_n-E_0} \right]^2
\,.
\end{multline}
We define the angular factors
\begin{align}
\widetilde{\beta}^{(N)}_\ell \equiv & \;
\begin{cases}
2 \beta_{0,\ell}^{(2)} \alpha_{\ell,1}^{(2)} \alpha_{1, 0}^{(2)} & \qquad (N=2)\,, \\
\beta_{0,\ell}^{(3)} \alpha_{\ell,1}^{(3)} \alpha_{1, 0}^{(3)} & \qquad (N=3) \,,
\end{cases} 
\\[0.1133ex]
\widetilde \gamma^{(N)} \equiv & \;
\begin{cases}
3 \beta_{1,1}^{(2)} \left( \alpha_{1,0}^{(2)}\right)^2 & \qquad (N=2) \,, \\
\beta_{1,1}^{(3)} \left( \alpha_{1,0}^{(3)}\right)^2 & \qquad (N=3) \,.
\end{cases}
\end{align}
Using the expressions given in 
Eqs.~\eqref{eq::alpha0}, \eqref{eq::alpha1}, \eqref{eq::beta0}, 
and~\eqref{eq::beta1} for the angular factors,
we finally get
\begin{multline}
\label{C21int}
\int_{-\infty}^{+\infty} \dd t_2\int_{-\infty}^{+\infty} \dd t_3 \,  
\widetilde{C}^{(1,2)}(t_2,t_3) = \\
4 \sum_{n, n'} \sum_{\ell=0,2} \widetilde \beta^{(N)}_\ell \frac{Q_{00, n \ell} 
S_{n \ell,n'1} S_{n'1,00}}{(E_n-E_0)(E_{n'}-E_{0})} \\
+ 2 \widetilde \gamma^{(N)} \sum_{n, n'} 
\frac{S_{00, n1} Q_{n1,n'1} S_{n'1,00}}{(E_n-E_0)(E_{n'}-E_{0})} 
\\
- 2 \widetilde \beta^{(N)}_0 \sum_{n} \frac{Q_{00,00} \, S^2_{00, n1} }{(E_{n}-E_0)^2} 
-4 \widetilde \alpha_0^{(N)} \left[ \sum_{n} \frac{S_{00, n1}^2}{E_n-E_0} \right]^2 
\\
- 8 \widetilde \alpha_0^{(N)} \sum_{n, n'} \frac{S^2_{00, n1} 
S^{2}_{00,n'1}}{(E_{n'}-E_0)(E_{n'}+E_{n}-2E_{0})} 
\\
\sim 
\frac{1}{N} \, \sum_K [G^{(1,2)}(N,1)]_K \, g^K \,.
\end{multline}
Results for perturbative coefficients $[G^{(1,2)}(N,1)]_K$
for the cases $N=2$ and $N=3$, for low orders 
of perturbation theory, are given in Tables~\ref{table2}
and~\ref{table3}, respectively.
This concludes our discussion of the formalism used
for obtaining higher orders of perturbation 
theory for the correlation functions discussed
in this articles.
We can now proceed to the comparison with the 
analytic large-order estimates, and the subleading 
corrections, evaluated in Ref.~\cite{GiEtAl2020}.

\begin{figure*}[t!]
\begin{center}
\includegraphics[width=0.8\linewidth]{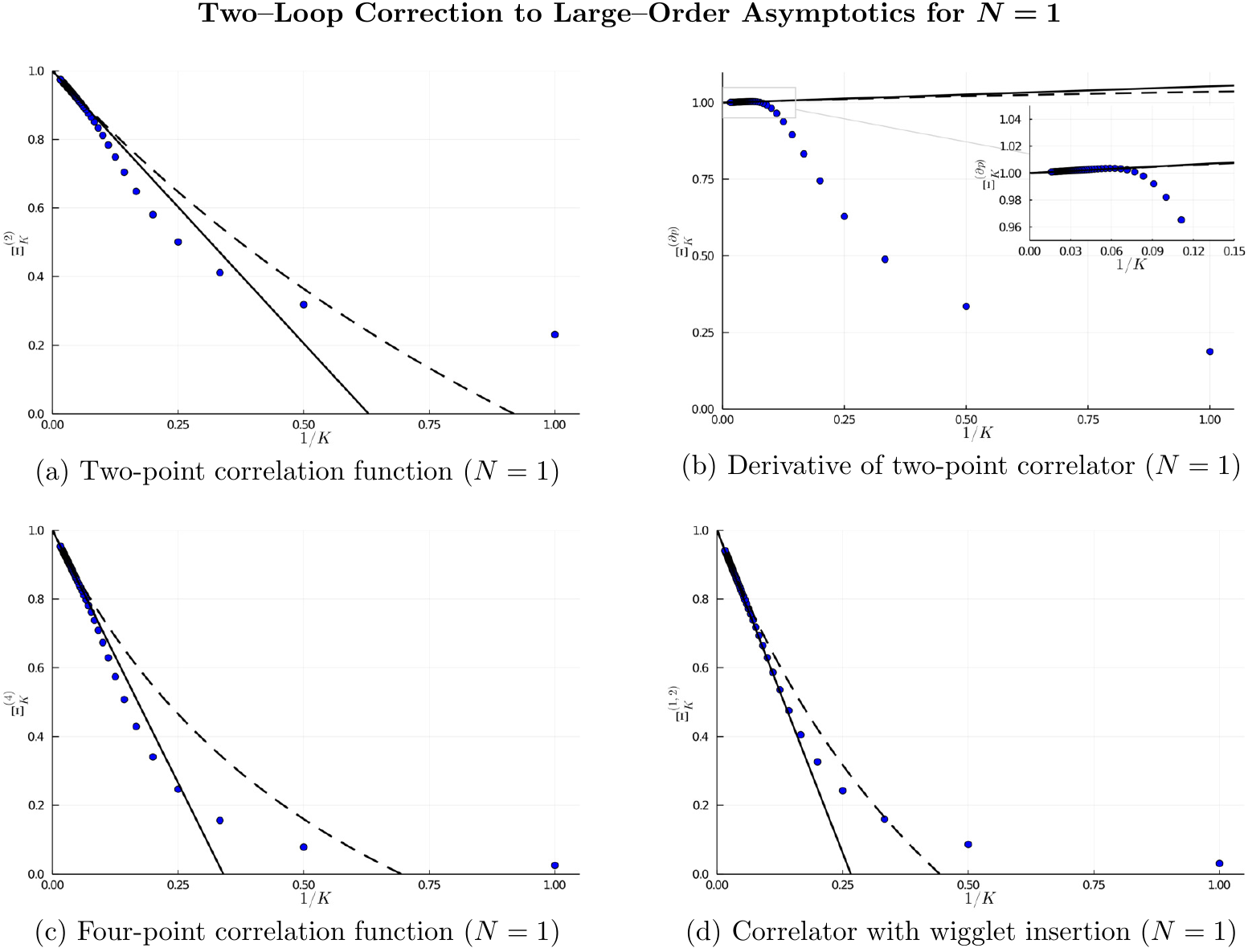}
\caption{\label{Fig2}Same as Fig.~\ref{Fig1}, but the correlation functions in $N=1$.}
\end{center}
\end{figure*}

\begin{figure*}[t!]
\begin{center}
\includegraphics[width=0.8\linewidth]{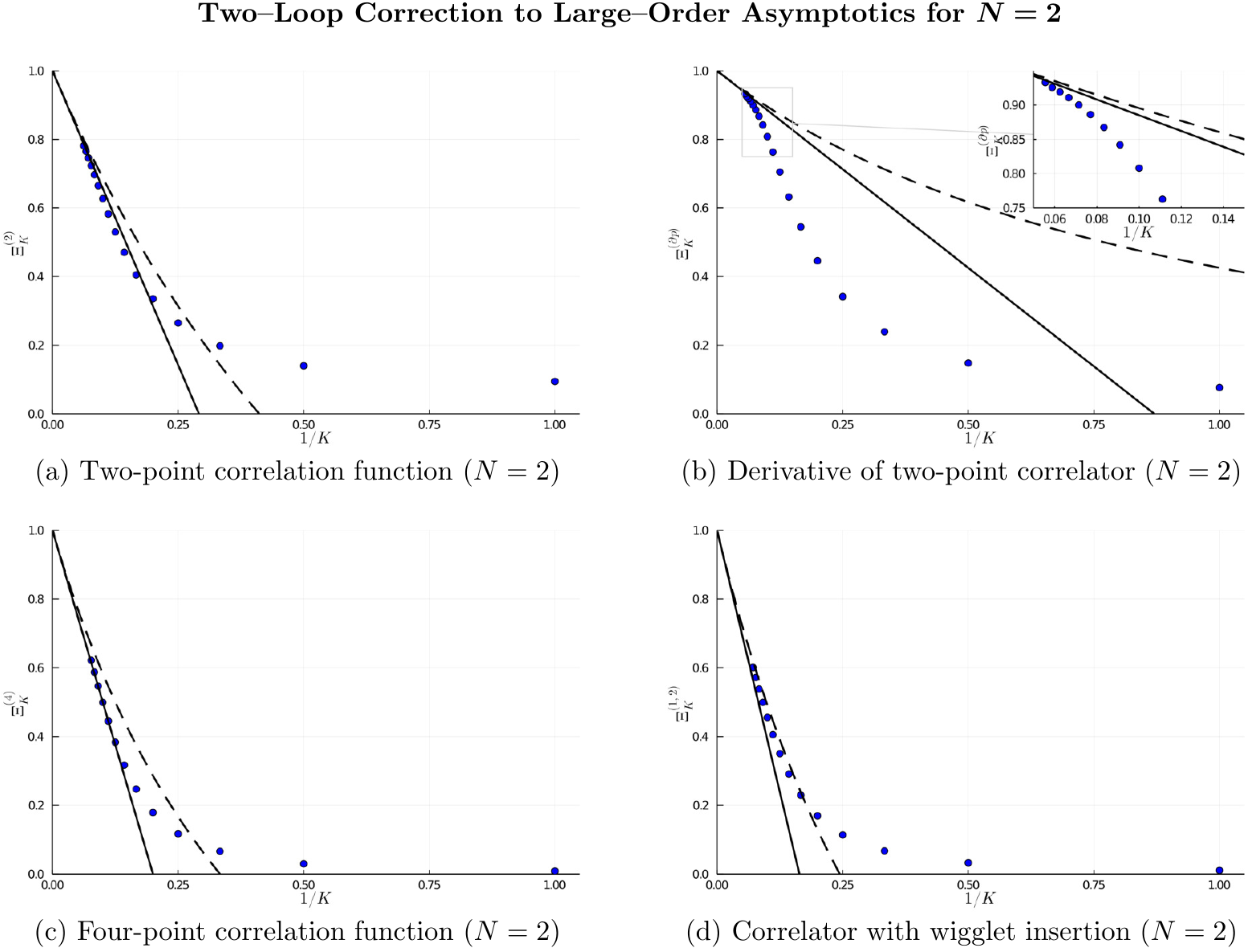}
\caption{\label{Fig3}Same as Fig.~\ref{Fig2}, but for $N=2$.}
\end{center}
\end{figure*}

\begin{figure*}[t!]
\begin{center}
\includegraphics[width=0.8\linewidth]{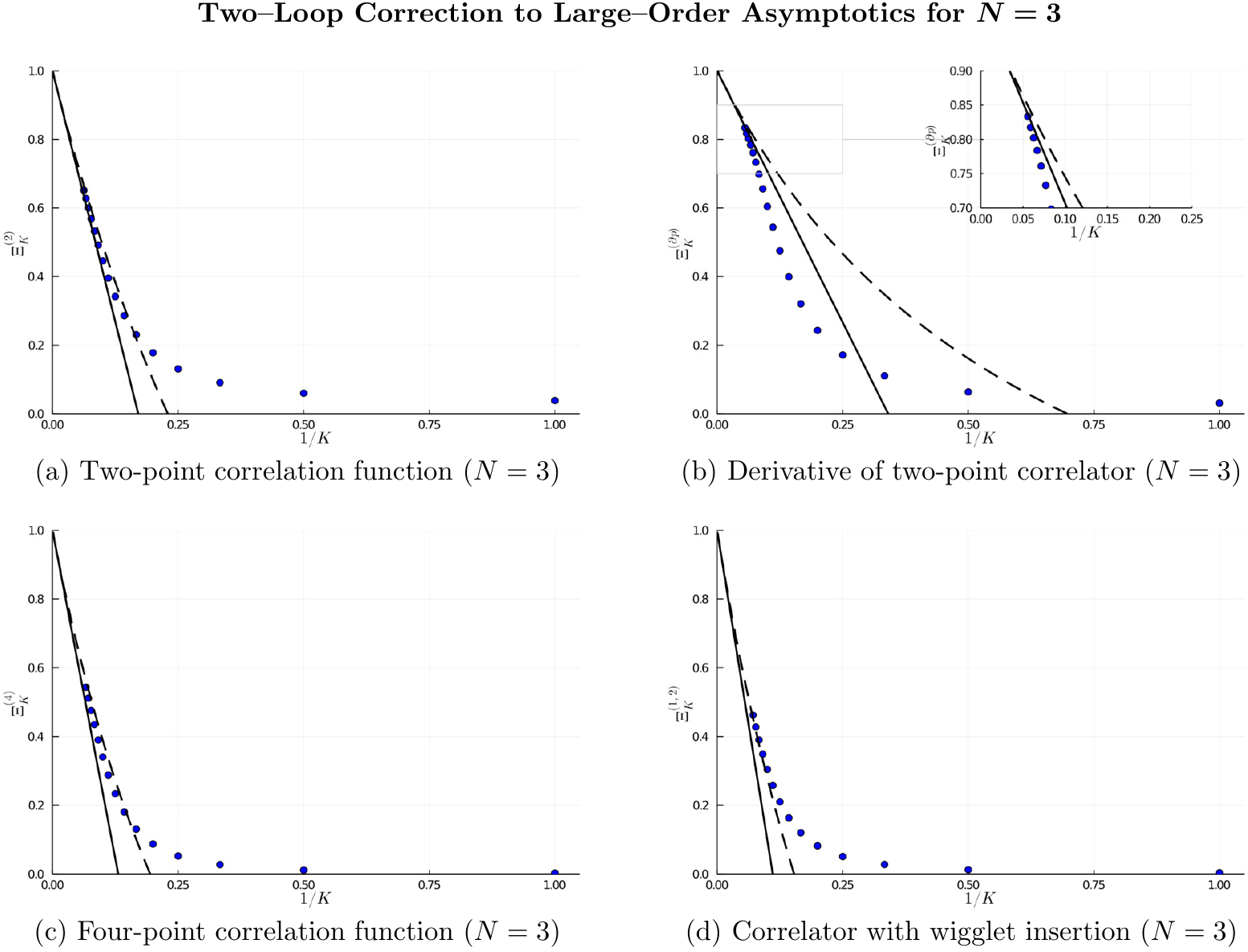}
\caption{\label{Fig4}Same as Fig.~\ref{Fig2}, but for $N=3$.}
\end{center}
\end{figure*}

%
%
\section{Comparison with Two--Loop Corrections for Large Order}
\label{sec4} 

\subsection{Analytic Formulas}
\label{sec4A}

In this section,
we briefly review the results obtained in~\cite{GiEtAl2020}
concerning the large order behaviour of the ground state energy and of an $M$-point correlation function $G$
for a $D$-dimensional field theory with $N$ components,
which can be traced to the two-loop corrections
to the instanton configurations that describe
the leading factorial growth of the perturbative
coefficients..
We refer to the
perturbative coefficient of order $K$ as $G_K^{(M)}$.  When $K$ is large, we can
express $G_K^{(M)}$ to order $1/K$ as
\begin{multline}
\label{genexp}
G_K^{(M)} = \frac{c(N, D)}{\pi} \, 
\Gamma \left(K+b\right)\left( \frac{1}{A} \right)^{b} \; 
\left(-\frac{1}{A}\right)^K \;
\left[ 1  \right.
\\[0.1133ex]
\left.
- \frac{A \, d(N, D)}{K + b - 1} + 
\calO(K^{-2}) \right] \,,
\qquad
b=\frac{M + N + D - 1}{2}\,,
\end{multline}
where $A=4/3$ is the action of the $\phi^4$ theory evaluated on the instanton
saddle-point multiplied by the coupling constant and with an inverted sign. For
large $K$, we can replace $K - 1 + b \to K$ in the denominator of the second
term and identify the $1/K$-correction. 

We start with the ground-state energy obtained using the relation in
Eq.~\eqref{C2int} computing the bracket of the 
perturbative Hamiltonian with the unperturbed and 
perturbed ground state eigenfunctions.
Specifically, one needs to examine the relation
\begin{equation}
\int_{-\infty}^{+\infty} C^{(0)}(t) \, \dd t \sim
\sum_K [G^{(0)}(N,1)]_K \, g^K \,.
\end{equation}
Here, $[G^{(0)}_{N,1}]_K$
is the specialization of the general asymptotic
perturbative coefficient $G_K$ of order $K$ 
given in Eq.~\eqref{genexp} to the ground-state energy
function ($M=0$) in one spatial dimension ($D=1$) 
and for an internal $O(N)$ symmetry group,
\begin{multline}
\label{G0N1K}
[G^{(0)}(N,1)]_K =
\frac{c^{(0)}(N, 1)}{\pi} \,
\Gamma \left(K+b\right)\left( \frac{1}{A} \right)^{b} \;
\left(-\frac{1}{A}\right)^K \;
\\[0.1133ex]
\times \left[ 1  
- \frac{A \, d^{(0)}(N,1)}{K + b - 1} +
\calO(K^{-2}) \right] \,,
\qquad
b= \frac{N}{2}\,,
\end{multline}
where we defined
\begin{subequations}
\label{c0pert}
\begin{align}
c^{(0)}(N,1) =& \;
2\pi^2\frac{8^{N/2}}{\Gamma(N/2)},\\
d^{(0)}(N,1) =& \; 
\frac{5}{24} + \frac{5}{2 \pi^2} - \frac{7 \zeta(3)}{4 \pi^2} 
\nonumber\\[0.1133ex]
& \; 
+ \left( \frac{9}{16} - \frac{1}{2 \pi^2} + \frac{7 \zeta(3)}{4 \pi^2} \right) N
+ \frac{7}{32}  \, N^2 \,.
\end{align}
\end{subequations}

For the two point correlation function given in 
Eq.~\eqref{C2int} one needs to examine the relation
\begin{equation}
\label{possible_contradiction}
\int_{-\infty}^{+\infty} C^{(2)}(t) \, \dd t \sim
\sum_K [G^{(2)}(N,1)]_K \, g^K \,.
\end{equation}
Here, $[G^{(2)}_{N,1}]_K$
is the specialization of the general asymptotic
perturbative coefficient $G_K$ of order $K$ 
given in Eq.~\eqref{genexp} to the two-point correlation
function ($M=2$) in one spatial dimension ($D=1$) 
and for an internal $O(N)$ symmetry group,
\begin{multline}
\label{G2N1K}
[G^{(2)}(N,1)]_K =
\frac{c^{(2)}(N, 1)}{\pi} \,
\Gamma \left(K+b\right)\left( \frac{1}{A} \right)^{b} \;
\left(-\frac{1}{A}\right)^K \;
\\[0.1133ex]
\times \left[ 1  
- \frac{A \, d^{(2)}(N,1)}{K + b - 1} +
\calO(K^{-2}) \right] \,,
\qquad
b= 1 + \frac{N}{2}\,,
\end{multline}
where we defined
\begin{subequations}
\label{c2pertdef1}
\begin{align}
c^{(2)}(N,1) =& \;
2\pi^2\frac{8^{N/2}}{\Gamma(N/2)},\\
d^{(2)}(N,1) =& \; 
\frac{5}{24} + \frac{5}{2 \pi^2} - \frac{7 \zeta(3)}{4 \pi^2} 
\nonumber\\[0.1133ex]
& \; 
+ \left( \frac{9}{16} - \frac{1}{2 \pi^2} + \frac{7 \zeta(3)}{4 \pi^2} \right) N
+ \frac{7}{32}  \, N^2 \,.
\end{align}
\end{subequations}
%

For the second derivative of the two-point 
correlation function at zero momentum,
the asymptotic relationship is 
given in Eqs.~\eqref{C2derivint},
\begin{equation}
\label{possible_contradiction2}
\int_{-\infty}^{+\infty} t^2\,C^{(2)}(t) \, \dd t \sim
\sum_K [G^{(\partial p)}(N,1)]_K \, g^K \,,
\end{equation}
where
\begin{multline}
\label{GderivN1K}
[G^{(\partial p)}(N,1)]_K =
\frac{c^{(\partial p)}(N, 1)}{\pi} \,
\Gamma \left(K+b\right)\left( \frac{1}{A} \right)^{b} \;
\\
\times 
\left(-\frac{1}{A}\right)^K \,
\left[ 1
- \frac{A \, d^{(\partial p)}(N,1)}{K + b - 1} +
\calO(K^{-2}) \right] \,,
\\
b= 1 + \frac{N}{2}\,,
\end{multline}
with
\begin{subequations}
\label{c2derivpert}
\begin{align}
c^{(\partial p)}(N,1) =& \;
-\pi^4\frac{8^{N/2}}{\Gamma(N/2)},\\
d^{(\partial p)}(N,1) =& \;
\frac{5}{24} + \frac{4}{\pi^4} 
- \frac{21 \, \zeta(3)}{\pi^4} 
- \frac{93 \, \zeta(5)}{2 \pi^4} 
\\[0.1133ex]
& \;
+ \left( -\frac{3}{16} - \frac{6}{\pi^4} + \frac{93 \zeta(5)}{2 \pi^4} \right) \, N
+ \frac{7}{32}  \, N^2 \,.
\nonumber
\end{align}
\end{subequations}
For the four-point correlation function, 
the relationship is, from Eqs.~\eqref{C4int},
\begin{multline}
\label{possible_contradiction3}
\int_{-\infty}^{+\infty} \dd t_2
\int_{-\infty}^{+\infty} \dd t_3
\int_{-\infty}^{+\infty} \dd t_4 \, C^{(4)}(t_2,t_3,t_4)
\\
\sim 
\sum_K [G^{(4)}(N,1)]_K \, g^K \,,
\end{multline}
where
\begin{multline}
\label{G4K}
[G^{(4)}(N,1)]_K = \frac{c^{(4)}(N, 1)}{\pi} \,
\Gamma \left(K+b\right)\left( \frac{1}{A} \right)^{b} \;
\left(-\frac{1}{A}\right)^K \;
\\[0.1133ex]
\times \left[ 1 
- \frac{A \, d^{(4)}(N, 1)}{K + b - 1} +
\calO(K^{-2}) \right] \,,
\qquad
b= 2 + \frac{N}{2}\,,
\end{multline}
with
\begin{subequations}
\label{c4derivpert}
\begin{align}
c^{(4)}(N,1) =& \;
4\pi^4\frac{8^{N/2}}{\Gamma(N/2)},\\
d^{(4)}(N,1) =& \; 
\frac{5}{24} + \frac{13}{\pi^2} 
- \frac{7 \, \zeta(3)}{2 \pi^2} 
\\[0.1133ex]
& \; 
+ \left( \frac{9}{16} - \frac{1}{\pi^2} + \frac{7 \zeta(3)}{2 \pi^2} \right) \, N
+ \frac{7}{32}  \, N^2 \,.
\nonumber
\end{align}
\end{subequations}

For the two-point correlation function
with a wigglet insertion, 
the relationship is, from Eqs.~\eqref{C21int},
\begin{multline}
\int_{-\infty}^{+\infty} \dd t_2\int_{-\infty}^{+\infty} \dd t_3 \,  
C^{(1,2)}(t_2,t_3)
\\
\sim 
\sum_K [G^{(1,2)}(N,1)]_K \, g^K \,,
\end{multline}
where
\begin{multline}
\label{G21K}
G^{(1,2)}_K = \frac{c^{(1,2)}(N, 1)}{\pi} \,
\Gamma \left(K+b\right)\left( \frac{1}{A} \right)^{b} \;
\left(-\frac{1}{A}\right)^K \;
\\[0.1133ex]
\times \left[ 1 
- \frac{A \, d^{(1,2)}(N, 1)}{K + b - 1} +
\calO(K^{-2}) \right] \,,
\qquad
b= 2 + \frac{N}{2}\,,
\end{multline}
with
\begin{subequations}
\label{c21derivpert}
\begin{align}
c^{(1,2)}(N, 1) =& \;
8\pi^2\frac{8^{N/2}}{\Gamma(N/2)}, \\
d^{(1,2)}(N, 1) = & \;
\frac{35}{24} + \frac{5}{2 \pi^2} 
- \frac{7 \, \zeta(3)}{4 \pi^2} 
\\[0.1133ex]
& \;
+ \left( \frac{15}{16} - \frac{1}{2 \pi^2} + \frac{7 \zeta(3)}{4 \pi^2} \right) \, N
+ \frac{7}{32}  \, N^2 \,.
\nonumber
\end{align}
\end{subequations}

For reference,
the first few perturbative coefficients
$[G^{(2)}(N,1)]_K$, $[G^{(\partial p)}(N,1)]_K$, $[G^{(4)}(N,1)]_K$, and
$[G^{(1,2)}(N,1)]_K$, for $K = 0,1,2,3,4,5$ and $10$,
for the internal $O(N)$ symmetry group,
with $N = 1,2,3$,
are also given in Tables~\ref{table1}---\ref{table3}.

\color{black} 

\subsection{Significance of the Two--Loop Correction}
\label{sec4B}

In Figs.~\ref{Fig1}---\ref{Fig4}, 
we plot the asymptotic expression of the
coefficients of the perturbative expansion of the ground-state energy and of the correlation functions as a
function of the inverse of the order of perturbation $K$. These coefficients have
been divided by their leading order expression reported in Eq.~\eqref{genexp},
that is the expression proportional to 
\begin{equation}
[Q^{(M)}]_K =\frac{c(N, 1)}{\pi} \, \Gamma \left(K+b\right)\left( \frac{1}{A}
\right)^{(M+N)/2} \; \left(-\frac{1}{A}\right)^K  \,.
\end{equation}
and they read
\begin{equation}
[\Xi^{(M)}]_K =\frac{[G^{(M)}]_K}{[Q^{(M)}]_K}.
\label{XiM}
\end{equation}
These coefficients have been compared
with their next-to-leading order estimate, i.e., 
with the multiplicative term
\begin{equation}
[\Xi^{(M)}]_K \approx
\mu_K^{(M)}=1-A \frac{d(M,N)}{K} \,,
\label{mu}
\end{equation}
and with 
\begin{multline}
[\Xi^{(M)}]_K \approx
\nu_K^{(M)} = 1-A \frac{d(M,N)}{K+b-1} \\
= 1 - A \frac{d(M,N)}{K} 
+ A \frac{(b-1)d(M,N)}{K^2}+O\left(\frac{1}{K}\right)^3 \,,
\label{nu}
\end{multline}
which carries a 
(part of the) next-to-next-to-leading $1/K^2$ correction term.

In Figs.~\ref{Fig1}--\ref{Fig4}, 
we observe a good agreement between the asymptotic estimate of
the perturbative coefficients of the correlation functions obtained 
in Ref.~\cite{GiEtAl2020} with the explicit higher-order calculations 
reported here.
Indeed, the improvement of the 
agreement upon the inclusion of the next-to-leading order
correction is quite remarkable.
The calculation of the large-order behavior of the
perturbative expansion of the correlation functions for $N>1$ was much more
computationally expensive with respect to the case $N=1$, therefore it was
possible to obtain less orders of perturbation.

%
%
\section{Conclusions}
\label{sec5}

In this article, we have discussed the explicit higher-order
calculation of the perturbative expansions of 
correlation functions for the $O(N)$ quartic anharmonic oscillator. 
We discussed the $N=1$ quantum anharmonic 
oscillator in Sec.~\ref{sec2}.
In Sec.~\ref{sec3}, we discussed the formulation of the 
perturbative expansion of the correlation functions of the 
$O(N)$ quantum anharmonic oscillator, 
where the internal symmetry group is assumed to be 
$O(2)$ or $O(3)$, and general formulas are given 
which allow us to enter a unified evaluation 
of the perturbative expansions.
Specifically, we considered the two-point
correlation function in Sec.~\ref{sec3B},
the four-point correlator in Sec.~\ref{sec3C},
and the correlation function with a wigglet 
insertion in Sec.~\ref{sec3D}.
The comparison with analytic results together with a review of the previously (Ref.~\cite{GiEtAl2020}) obtained
results for the large-order behavior of the 
correlation functions was
carried out in Sec.~\ref{sec4}.
The data in Figs.~\ref{Fig1}---\ref{Fig3}
underline the importance of the next-to-leading order
correction to the large-order factorial 
growth of the perturbative coefficients
for the demonstration of the agreement
of asymptotic estimates and explicit perturbative 
calculations.

Let us take, as an example, the coefficient
of order $g^8$ (the ``eight-loop coefficient'') 
for the two-point correlation function in the $O(3)$ model.
The explicit result is
\begin{equation}
[G^{(4)}_{N=3}]_{K=8} \simeq 3.03 \cdot 10^6 \,.
\end{equation}
The leading asymptotic term is
\begin{equation}
[G^{(4)}_{N=3,1}]_{K=8} \approx 8.87 \cdot 10^6 \,.
\end{equation}
With the inclusion of the two-loop (order $1/K$) correction, 
we find the (much better) estimate
\begin{equation}
[G^{(4)}_{N=3,1}]_{K=8} \approx 2.38 \cdot 10^6 \,.
\end{equation}
which differs from the the exact perturbative 
coefficient by roughly $20$\,percent.
At order $K=10$, we already observe
$93$\,percent agreement, whereas at order $K=14$,
the agreement it is slightly better than $97$ percent. 
In some other cases, the agreement is
surprisingly good even at very low orders. 
For example, for $N=2$, the two-loop 
large-order estimate of the coefficient of the 
four-point correlation function agree
with the exact perturbative coefficient
at the level of 98\,percent,
in eight-loop order ($K=8$).

The tests presented here are essential to have a good starting point from which
to extend the calculations to field theory, i.e., to the case $D>1$, where this
type of checks are not possible anymore.  Specifically, the agreement between
these two different approaches ensures the correctness of the method described
in \cite{GiEtAl2020}, which can be then generalized to obtain the perturbative
expression of the correlation functions for two and three dimensional
$N$-vector model where is not possible to use the conventional techniques of
perturbation theory.  We recall that, irrespective of the dimension $D$,  the
same dispersion relation relates the large-order growth of the coefficients of
a correlation function with the behaviour at small orders of perturbation of
its imaginary part for negative coupling.

\section*{Acknowledgements}
This work has been supported by the National Science Foundation 
(Grant No.~PHY--2110294), 
by the Swedish Research Council (Grant No.~638--2013--9243) 
and by the Simons Foundation (Grant No.~454949).

\appendix 

%
%
\section{Some useful definitions and identities} 
\label{app:identities}

%
%
\subsection{Matrix elements}
\label{app:identitiesA}

In this section, we will derive the expression for the 
various angular matrix elements denoted as $\alpha$,
and $\beta$, which appear in the main text.  We start considering the matrix element
$\langle n, \vec{\eta}|r \cos \theta_1| n', \vec{\eta}'\rangle$.
If the coordinate $r \cos \theta_1$ is aligned
with the quantization axis, then the only nonvanishing 
transition matrix element will be obtained 
for all magnetic projections equal to zero.
We can therfore assume that 
$\ket{n, \vec{\eta}} \equiv \ket{n, \ell, 0, \dots, 0}$ and
$\ket{n', \vec{\eta}'} \equiv \ket{n', \ell', 0, \dots, 0}$. 
Under these assumptions, we can write the
matrix element as
\begin{subequations}
\label{matrix_element1}
\begin{equation}
\langle n, \vec{\eta} |r \cos \theta_1| n', \vec{\eta}'\rangle = 
S_{n \ell, n' \ell'} \, \alpha^{(N)}_{\ell, \ell'} 
\end{equation}
where the radial part is given as follows,
\begin{align}
S_{n\ell, n'\ell'} &\equiv \int dr \, r^N R_{n \ell}(r) R_{n' \ell'}(r) \\
\alpha^{(N)}_{\ell, \ell'} &\equiv \int \dd \theta_1 \dots \dd \theta_{N-1} \, 
\cos(\theta_1) J(\theta_1, \dots, \theta_{N-1}) 
\nonumber\\[0.1133ex]
& \; \times Y^{0 \dots 0*}_{\ell}(\theta_1, \dots, \theta_{N-1}) \,
Y^{0 \dots 0}_{\ell'} (\theta_1, \dots, \theta_{N-1}) \,,
\end{align}
\end{subequations}
and $J(\theta_1, \dots, \theta_{N-1})$ is the 
Jacobian due to the hyperspherical change of coordinates. 

Due to the orthogonality relations between the hyperspherical 
harmonics, only few of the $\alpha^{(N)}_{\ell, \ell'}$ 
terms are non-zero. For example, if $\ell=0$,
then we have for $N=2$ and $N=3$ respectively,
\begin{subequations}
\label{eq::alpha0}
\begin{align}
\alpha^{(2)}_{0,\ell'} &= \frac{1}{2} \left( \delta_{\ell', 1} 
+ \delta_{\ell', -1} \right) \,, \\
\alpha^{(3)}_{0,\ell'} &= \frac{1}{\sqrt{3}} \delta_{\ell', 1} \,.
\end{align}
\end{subequations}
When $\ell=1$, one instead obtains
\begin{subequations}
\label{eq::alpha1}
\begin{align}
\alpha^{(2)}_{1,\ell'} &= \frac{1}{2} 
\left( \delta_{\ell', 0} + \delta_{\ell', 2} \right) \,, \\
\alpha^{(3)}_{1,\ell'} &= \frac{1}{\sqrt{3}} \delta_{\ell', 0} + 
\frac{2}{\sqrt{15}} \delta_{\ell', 2} \,.
\end{align}
\end{subequations}
In general, one has
\begin{equation}
\alpha_{\ell, \ell'}^{(2)} = \frac{1}{2}\left( \delta_{\ell', \ell+1} + \delta_{\ell', \ell-1}\right).
\end{equation}

For the two-point function with a wigglet insertion,
we also have to consider the matrix element 
$\langle n, \vec{\eta}|(r \cos \theta_1)^2| n',\vec{\eta}'\rangle$ 
where $\ket{n, \vec{\eta}} \equiv
\ket{n, \ell, 0, \dots, 0}$ and 
$\ket{n', \vec{\eta}'} \equiv \ket{n', \ell', 0, \dots, 0}$. 
It can be written as
\begin{subequations}
\label{matrix_element2}
\begin{equation}
\langle n, \vec{\eta}|(r \cos \theta_1)^2| n', \vec{\eta}'\rangle = 
Q_{n \ell, n' \ell'} \, \beta^{(N)}_{\ell, \ell'}  \,,
\end{equation}
where
\begin{align}
Q_{n \ell, n' \ell'} \equiv & \; \int \dd r \, r^{N+1} \, R_{n \ell}(r) \, R_{n' \ell'}(r)  \,,
\\[0.1133ex]
\beta^{(N)}_{\ell, \ell'} \equiv & \; \int \dd \theta_1 \dots \dd \theta_{N-1} \, 
(\cos(\theta_1))^2 J(\theta_1, \dots, \theta_{N-1}) 
\nonumber\\[0.1133ex]
& \; \times  Y^{0 \dots 0*}_{\ell} (\theta_1, \dots, \theta_{N-1}) \;
Y^{0 \dots 0}_{\ell'} (\theta_1, \dots, \theta_{N-1}) \,.
\end{align}
\end{subequations}
Also in this case, because of the orthogonality relations between the
hyperspherical harmonics, only few of the 
$\beta^{(N)}_{\ell, \ell'}$ terms are
non-zero. For example, if $\ell=0$,
then we have for $N=2$ and $N=3$, respectively
\begin{subequations}
\label{eq::beta0}
\begin{align}
\beta^{(2)}_{0, \ell'} &= \frac{1}{4} \left( 2\delta_{\ell', 0} + 
\delta_{\ell', 2}+\delta_{\ell',-2} \right) \,, 
\\ 
\beta^{(3)}_{0,\ell'} &= 
\frac{1}{3}\left( \delta_{\ell', 0}+\frac{2}{\sqrt{5}}\delta_{\ell',2}\right) \,.
\end{align}
\end{subequations}
When $\ell=1$, instead, one has
\begin{subequations}
\label{eq::beta1}
\begin{align}
\beta^{(2)}_{1,\ell'} =& \; \frac{1}{4} \left( \delta_{\ell', -1} + 2\delta_{\ell', 1} 
+ \delta_{\ell', 3}\right) \,, \\ 
\beta^{(2)}_{-1,\ell'} =& \; \frac{1}{4} \left( 2\delta_{\ell', -1} + 
\delta_{\ell', 1} +\delta_{\ell', -3}\right) \, \\
\beta^{(3)}_{1,\ell'} =& \; \frac{1}{5} \left(3 \, \delta_{\ell', 1} + 
2 \sqrt{\frac{3}{7}} \, \delta_{\ell', 3}\right) \,.
\end{align}
\end{subequations}
The above formulas can be used to perform all 
required angular integrals for the correlation
functions considered in our investigations.

%
%
\subsection{Case $\maybebm{N=3}$}
\label{app:identitiesB}

For $N=3$, the coefficients 
$\alpha_{\ell,\ell'}^{(N=3)}$ and 
$\beta_{\ell,\ell'}^{(N=3)}$ can be
written in terms of the Gaunt coefficients 
$Y^{\ell \ell' \ell''}_{m m' m''}$
defined as the integral over three
spherical harmonics
\begin{equation}
\begin{split}
Y^{\ell \ell' \ell''}_{m m' m''} 
&= \int_0^{2 \pi} \dd \varphi \, 
\int_0^\pi \dd \theta 
\sin(\theta)\\
& \; \times Y^*_{\ell \, m}(\theta, \varphi) \,
Y_{\ell' m'}(\theta, \varphi)\,
Y_{\ell'' m''}(\theta, \varphi) \,.
\label{gaunt}
\end{split}
\end{equation}
By writing $\cos(\theta)$ and $\cos(\theta)^2$ in 
terms of spherical harmonics we get
\begin{equation}
\alpha^{(3)}_{\ell, \ell'} \equiv 2\sqrt{\frac{\pi}{3}}Y^{\ell 1 \ell'}_{0 0 0}  \,,
\end{equation}
and
\begin{equation}
\beta^{(3)}_{\ell, \ell'} \equiv 
\frac{1}{3}\left(4\sqrt{\frac{\pi}{5}} \, Y^{\ell 2 \ell'}_{0 0 0} +
2\sqrt{2}\, Y^{\ell 0 \ell'}_{0 0 0}\right) \,.
\end{equation}
Alternatively, 
the integral appearing in Eq. (\ref{gaunt}) can be interpreted using the
Wigner-Eckart theorem and can be written as the product of the Clebsch-Gordan
coefficient corresponding to the quantum numbers $\ell$, 
$\ell'$, $\ell''$, $m$, $m'$, and $m''$,
and the reduced matrix element of the spherical harmonic tensor
$Y^{m'}_{\ell'}$, as
\begin{multline}
\int_0^{2 \pi} \dd \varphi 
\int_0^\pi \dd \theta \sin(\theta) \,
Y^*_{\ell \, m}(\theta, \varphi) \,
Y_{\ell' \, m'}(\theta, \varphi)\,
Y_{\ell'' \, m''}(\theta, \varphi) \\
=
(-1)^{\ell'-\ell''+\ell}
\frac{C_{\ell'm'\ell''m''}^{\ell m}}%
{\sqrt{2\ell+1}}\langle\ell||\vec{Y}_{\ell'}||\ell''\rangle,
\end{multline}
where the reduced matrix can be expressed using a $3j$ symbol 
with three zero magnetic projections
\begin{multline}
\langle\ell||\vec{Y}_{\ell'}||\ell''\rangle =
(-1)^{\ell}\sqrt{\frac{(2\ell+1)(2\ell'+1)(2\ell''+1)}{4\pi}}
\\
\times 
\left( \begin{matrix}
\ell & \ell' & \ell''\\
0 & 0 & 0
\end{matrix}\right).
\end{multline}
Using this convention the coefficients
$\alpha_{\ell,\ell'}^{(3)}$ and $\beta_{\ell,\ell'}^{(3)}$ 
will read
\begin{equation}
\alpha^{(3)}_{\ell, \ell'} \equiv 
(-1)^{1-\ell'+\ell} \, 
2\sqrt{\frac{\pi}{3}} \, 
\frac{C_{10\ell'm'}^{\ell m}}{\sqrt{2\ell+1}} \,
\langle\ell||\vec{Y}_{1}||\ell'\rangle \,,
\end{equation}
and
\begin{multline}
\beta^{(3)}_{\ell, \ell'} \equiv 
\frac{1}{3} \, \bigg((-1)^{2-\ell'+\ell}
4\sqrt{\frac{\pi}{5}} \, 
\frac{C_{20\ell'm'}^{\ell m}}{\sqrt{2\ell+1}} \,
\langle\ell||\vec{Y}_{2}||\ell'\rangle \\
+(-1)^{-\ell'+\ell}2\sqrt{2}
\frac{C_{00\ell'm'}^{\ell m}}{\sqrt{2\ell+1}} \, 
\langle\ell||\vec{Y}_{0}||\ell'\rangle\bigg) \,.
\end{multline}
These results are in agreement with the formulas
obtained in Appendix~\ref{app:identitiesA}.

%
%
\section{Alternative Procedure}
\label{alternative}

The perturbative treatment of the radial part 
can be accomplished by a direct mapping of the 
procedure outlined in Eqs.~\eqref{unperturbed}---\eqref{direct}
onto a computer algebra system.
However, it is useful 
to delineate an alternative procedure to compute the eigenvalues
and eigenfunctions of the perturbed three dimensional harmonic oscillator. 
\color{black}
We will deal with the $N=3$ case; the 
generalization to $N=2$ is relatively straightforward. We
adapt to our case the results of Ref.~\cite{EhMe1972}, where the
computation has been carried out for a general central field perturbation. The
eigenvalues and eigenfunctions of our Schrödinger equation 
\begin{multline}
\left( -\vec\nabla^2 + r^2+ \frac{g}{2} r^4 \right)
\Psi_{n \ell m}(r, \theta,\varphi)
\\
= \alpha_{n \ell}\Psi_{n \ell m}(r, \theta,\varphi) \,,
\qquad \alpha_{n \ell} =  2E_{n \ell}\,,
\end{multline}
where $E_{n \ell}$
can be written as a perturbative series in the 
coupling parameter $g$,
\begin{subequations}
\begin{align}
\alpha_{n \ell} =& \; \sum_{K=0}^\infty 
\alpha_{n\ell, K} \, g^K \,, 
\\  
\Psi_{n \ell m}(r, \theta, \varphi) =& \;
R_{n \ell}(r) \, Y_{\ell \, m}(\theta,\varphi) \,,
\\  
R_{n \ell}(r) =& \;
\calN_{n \ell} \, \ee^{-r^2/2} \, r^\ell \,
\sum_{K=0}^\infty u_{n \ell}^{(K)}(r^2) \, g^K \,,
\end{align}
\end{subequations}
where $\calN_{n \ell}=\calN_{n \ell}^{(N=3)}$ is defined in Eq. (\ref{eq::norm3N}). Each coefficient $u_{n\ell}^i(r^2)$ can be expressed as a linear combination
of the eigenfunctions of the unperturbed case
(it can be shown that only $4K$
terms will contribute to the $K$th order perturbation)
\begin{subequations}
\begin{align}
u_{n \ell}^{(K)}(r^2) =& \;
\sum_{j=\max(q_{n \ell}-2K, 0)}^{q_{n \ell} +2K} 
A_{K j} \, L_j^{\ell+\frac{1}{2}}(r^2)
\\
q_{n \ell} =& \; \alpha_{n \ell, 0}-3-2 \ell =\frac{1}{2}(n- \ell).
\end{align}
\end{subequations}
The coefficients $\alpha_{n\ell,K}$ and $A_{K j}$ can be determined from 
Eqs.~(32) and~(33) of Ref.~\cite{EhMe1972} by setting 
$B_j=\frac{\delta_{1j}}{2}$. 
Special cases are
\begin{equation}
A_{0,j} = \delta_{j,q_{n \ell}}, \quad A_{K,q_{n \ell}}=0\; 
\forall K > 0 \,. 
\end{equation}
However, there are some typos 
in the passages of the paper and we report here a corrected version 
of the main passages needed to arrive to the result.
\color{black}
We use the following recursive relation of the generalized
Laguerre polynomials,

\begin{equation}
r^m L^{\ell+\frac{1}{2}}_j(r^2) =
\sum_{n=0}^{2m} a(m,n,j) \, L^{\ell+\frac{1}{2}}_{j+m-n}(r^2) \,,
\end{equation}
where
\begin{multline}
a(m,n,j)= \frac{(-1)^{m-n}(m!)^2}{\Gamma(j+m-n+\ell+\frac{3}{2})}
\\
\times \sum_{k=\max(0,j-n)}^{\min(j+m-n,j)}
\frac{\binom{j+m-n}{k} \, \Gamma(m+k+\ell+\frac{3}{2})}{(j-k)!(n-j+k)!(m-j+k)!}.   
\end{multline}
We can then rewrite Eq. (26) of \cite{EhMe1972} as
\begin{multline}
\sum_{j=0}^\infty 
(q_{n \ell}-j) A_{K j} \, L_j^{\ell+\frac{1}{2}}(r^2)
\\
= \frac{1}{8}
\sum_{w=0}^{K-1}\delta_{1(K-w)}
\sum_{j=0}^\infty A_{wj}\sum_{i=0}^{2(K-w+1)}
a(K-w+1,i,j) \, 
\\
\times L_{j+K-w+1-i}^{\ell+\frac{1}{2}}(r^2) - 
\frac{1}{4}\sum_{w=0}^{K-1}\alpha_{n\ell}^{(K-w)}
\sum_{j=0}^\infty A_{wj}L_j^{\ell+\frac{1}{2}}(r^2)
\\
= \frac{1}{8}
\sum_{j=0}^\infty A_{(K-1)j}
\sum_{i=j-2}^{j+2}
a(2,j+2-i,j) \, L_{i}^{\ell+\frac{1}{2}}(r^2) 
\\
- \frac{1}{4}\sum_{w=0}^{K-1} \alpha_{n \ell}^{(K-w)}
\sum_{j=0}^\infty A_{wj}L_j^{\ell+\frac{1}{2}}(r^2) \,.
\end{multline}
For a specific value of $L_s^{\ell+\frac{1}{2}}(r^2)$, 
with the convention
\begin{equation}
\Theta(0) = 1 
\end{equation}
for the Heaviside step function $\Theta$, we get
\begin{multline}
A_{K s} =\frac{1}{8(q_{n \ell}-s)}
\sum_{j=0}^\infty A_{(K-1)j} \, a(2,j+2-s,j) 
\\
\times \Theta(2-|s-j|)-\frac{1}{4(q_{n\ell}-s)}
\sum_{w=0}^{K-1}  \alpha_{n \ell}^{(K-w)}A_{w,s}
\\
=\frac{1}{8(q_{n \ell}-s)}
\sum_{\max( j=q_{n\ell}-2(K-1), 0)}^{q_{n\ell}+2(K-1)}
A_{(K-1)j} \, a(2,j+2-s,j)
\\
-\frac{1}{4(q_{n \ell}-s)}
\sum_{w=0}^{K-1} \alpha_{n \ell}^{(K-w)}
A_{w,s} \, \alpha_{n \ell, K}
\\
=\frac{1}{2}
\sum_{j= \max(q_{n \ell}-2(K-1), 0)}^{q_{n \ell} +
2(K-1)} \, A_{(K-1)j}a(2,j+2-q_{n \ell},j) \,.
\end{multline}
We can therefore write a recursive relation that will provide an expression for
the perturbative coefficients of the eigenvalues and eigenfunctions of the
Schrödinger equation. The two-dimensional case can be derived from
three-dimensional one using the expression for the eigenfunctions reported in
Eq.~\eqref{eigen2d}.

\end{document}